\documentclass[cup6b]{cupbook}

\usepackage{epsfig}

\newcommand{\gtrsim}{\lower.7ex\hbox{$\,\stackrel{\textstyle>}{\sim}\,$}}
\newcommand{\lesssim}{\lower.7ex\hbox{$\,\stackrel{\textstyle<}{\sim}\,$}}

\begin{document}

\setcounter{chapter}{10}

\author[J. Granot \& E. Ramirez-Ruiz]{Jonathan Granot$^1$ and Enrico
  Ramirez-Ruiz$^2$ \\ 
(1) Centre for Astrophysics Research, University
  of Hertfordshire, College Lane,\\
   Hatfield, Herts, AL10 9AB, UK \\ 
(2) Astronomy and Astrophysics Department, University of California,\\ 
    Santa Cruz, CA 95064, USA}

%\chapter{GRB Unification Schemes}

\chapter{Jets and Gamma-Ray Burst Unification Schemes}

\section*{Abstract}
\noindent
There are several lines of evidence indicating that the
ultra-relativistic outflows powering gamma-ray bursts (GRBs) are
collimated into narrow jets. However, these are indirect, and the jet
structure is rather poorly constrained. What is more, the jet dynamics
have still not been investigated in detail. It has been suggested that
the observed variety between different long duration events, ranging
from bright spectrally hard GRBs, to dimmer and spectrally softer
X-ray flashes (XRFs) may be largely due to different viewing angles
(or lines of sight) relative to rather similar relativistic jets. Here
we describe the current state of knowledge on these topics, explain
some of the most relevant physics behind some of the basic principles,
and discuss prospects for the future.

%\section{Introduction}

\section{Evidence for bulk relativistic motion in gamma-ray bursts}
\label{rel_motion}

The first line of evidence for ultra-relativistic bulk motion of the
outflows that produce GRBs arises from the compactness argument. It
relies on the observed short and intense pulses of gamma rays and
their non-thermal energy spectrum that often extends up to high photon
energies. Together, these facts imply that the emitting region must be
moving relativistically. In order to understand this better, let us
first consider a source that is either at rest or moves at a Newtonian
velocity, $\beta \equiv v/c \ll 1$, corresponding to a bulk Lorentz
factor $\Gamma \equiv (1-\beta^2)^{-1/2} \approx 1$. For such a source
the observed variability timescale (e.g., the width of the observed
pulses) $\Delta t$, implies a typical source size or radius $R <
c\Delta t$, due to light time travel effects (for simplicity we ignore
here cosmological effects, such as redshift or time dilation). GRBs
often show significant variability down to millisecond timescales,
implying $R< 3\times 10^{7}(\Delta t/\,{\rm 1\,ms})$ cm.  At
cosmological distances their isotropic equivalent luminosity, $L$, is
typically in the range of $10^{50}-10^{53}\,{\rm erg\,s^{-1}}$. In
addition, the (observed part of the) $\varepsilon F_\varepsilon$ GRB
spectrum typically peaks around a dimensionless photon energy of
$\varepsilon\equiv E_{\rm ph}/m_e c^2\sim 1$, so that (for a Newtonian
source) a good fraction of the total radiated energy is carried by
photons that can pair produce with other photons of similar energy.
($F$ is the radiative flux and $F_\varepsilon\equiv{\rm d}F/{\rm
d}\varepsilon$). A simple estimate of the opacity to pair production
($\gamma\gamma \to e^+e^-$) usually results in a huge optical depth
for this process, $\tau_{\gamma\gamma}(\varepsilon) \sim \sigma_{\rm
T}n_\gamma(1/\varepsilon) R \sim \sigma_{\rm T} L_{1/\varepsilon} /
4\pi m_e c^3 R \gtrsim \sigma_{\rm T} L_{1/\varepsilon} / 4\pi m_e
c^4\Delta t \sim 10^{14}(L_{1/\varepsilon}/10^{51}\,{\rm
erg\,s^{-1}})(1\,{\rm ms}/\Delta t)$, where $L_\varepsilon \equiv
dL/d\varepsilon$ and $\sigma_{\rm T}$ is the Thomson cross section
(Granot et~al.\ 2008).  Such huge optical depths are clearly
inconsistent with the non-thermal GRB spectrum, which has a
significant power-law high-energy tail. This is known as the
compactness problem (Ruderman 1975).

If the source is moving relativistically toward us with a bulk Lorentz
factor $\Gamma \gg 1$, then in its own rest frame (where quantities
measured in that rest frame are denoted by a prime) the photons have
much smaller energies, $\varepsilon' \sim \varepsilon/\Gamma$, while
in the lab frame (i.e., the rest frame of the central source) most of
the photons propagate at angles $\lesssim 1/\Gamma$ relative to its
direction of motion (see Sect.~\ref{beaming} on aberration of light). The
latter implies that in the lab frame the typical angle between the
directions of the interacting photons is $\theta_{12} \sim 1/\Gamma$,
which has the following consequences. First, it increases the
threshold for pair production,
$\varepsilon_1\varepsilon_2>2/(1-\cos\theta_{12})$, to
$\varepsilon_1\varepsilon_2 \gtrsim \Gamma^2$ (compared to
$\varepsilon'_1\varepsilon'_2 \gtrsim 1$ for the roughly isotropic
distribution of angles between the directions of the interacting
photons in the rest frame of the source, where $\theta'_{12}\sim 1$).
Thus, $L_{1/\varepsilon}$ needs to be replaced by
$L_{\Gamma^2/\varepsilon}=\Gamma^{2(1-\alpha)}L_{1/\varepsilon}$,
where $L_\varepsilon\approx L_0\varepsilon^{1-\alpha}$ at high photon
energies (corresponding to $dN_{\rm ph}/d\varepsilon \propto
\varepsilon^{-\alpha}$, i.e., $\alpha$ is the high-energy photon
index). This results in an additional factor of
$\Gamma^{2(1-\alpha)}$ in the expression for
$\tau_{\gamma\gamma}(\varepsilon)$. Second, the expression for the
optical depth includes a factor of $1-\cos\theta_{12}$ (that
represents the rate at which photons pass each other and have an
opportunity to interact) which for a stationary source is $\sim 1$,
but for a relativistic source  moving toward us is $\sim
\Gamma^{-2}$.  Finally, the emission radius can be as large as $R
\lesssim \Gamma^2 c\Delta t$, which introduces an additional factor of
$\sim \Gamma^{-2}$ in the expression for $\tau_{\gamma\gamma}$.
Altogether, $\tau_{\gamma\gamma}(\varepsilon)$ is reduced by a factor
of $\sim\Gamma^{2(\alpha+1)}$
(i.e., $\tau_{\gamma\gamma}\propto\Gamma^{-2(\alpha+1)}$). Since
typically $\alpha\sim 2-3$, this usually implies $\Gamma \gtrsim 10^2$
in order to have $\tau_{\gamma\gamma} < 1$ and overcome the
compactness problem. Using similar arguments, the lack of a
high-energy cutoff due to pair production in the observed spectrum of
the prompt $\gamma$-ray emission in GRBs has been used to place lower
limits on the Lorentz factor of the outflow (Krolik \& Pier 1991,
Fenimore et~al.\ 1993, Woods \& Loeb 1995, Baring \& Harding
1997, Lithwick \& Sari 2001). Recently, high-energy observations by
the Fermi Gamma-Ray Space Telescope have enabled particularly strict
($\Gamma\gtrsim 10^3$) limits to be placed for both long (Abdo et
al. 2009a,b) and short (Ackermann et~al.\ 2010) GRBs.

A different, and somewhat complementary, line of evidence for
relativistic bulk motion in GRBs comes from estimates of the afterglow
image size at relatively late times (weeks to years) in the radio.
The afterglow image size increases with time, and therefore it can be
(marginally) angularly resolved only at late times, and only for
relatively nearby and radio bright afterglows. At such late times the
flow is much less relativistic, with a much more modest Lorentz
factor. The size of the afterglow image at a single epoch can be
estimated from the quenching of diffractive scintillation in the
radio afterglow (Goodman 1997), as the angular size of the source
becomes larger than that of the relevant density fluctuations in the
interstellar medium of our Galaxy, which is estimated to be roughly
$\theta_d \sim 3\,\mu$as. When the source angular size $\theta_{\rm s}$ is
smaller than $\theta_d$ then $\sim (\theta_d/\theta_{\rm s})^2
\gg 1$, and different sub-images are formed, which produce a random
diffraction pattern resulting in frequency dependent brightness
fluctuations of order unity ($\Delta F/F \sim 1$). For GRB\,970508 this
implied an apparent source size of $\sim 10^{17}\,$cm at $t \sim
30\,$days, or an average apparent expansion velocity close to the
speed of light, $c$, during the first month (Frail et~al.\ 1997, Taylor
et~al.\ 1997, Waxman et~al.\ 1998). The flux below the
synchrotron self-absorption frequency can also be used to constrain
the size of the emitting region (e.g., Katz \& Piran 1997, Granot et~al.\
2005).

\begin{figure*}
%\begin{center}
\epsfxsize=5.1cm \epsfbox{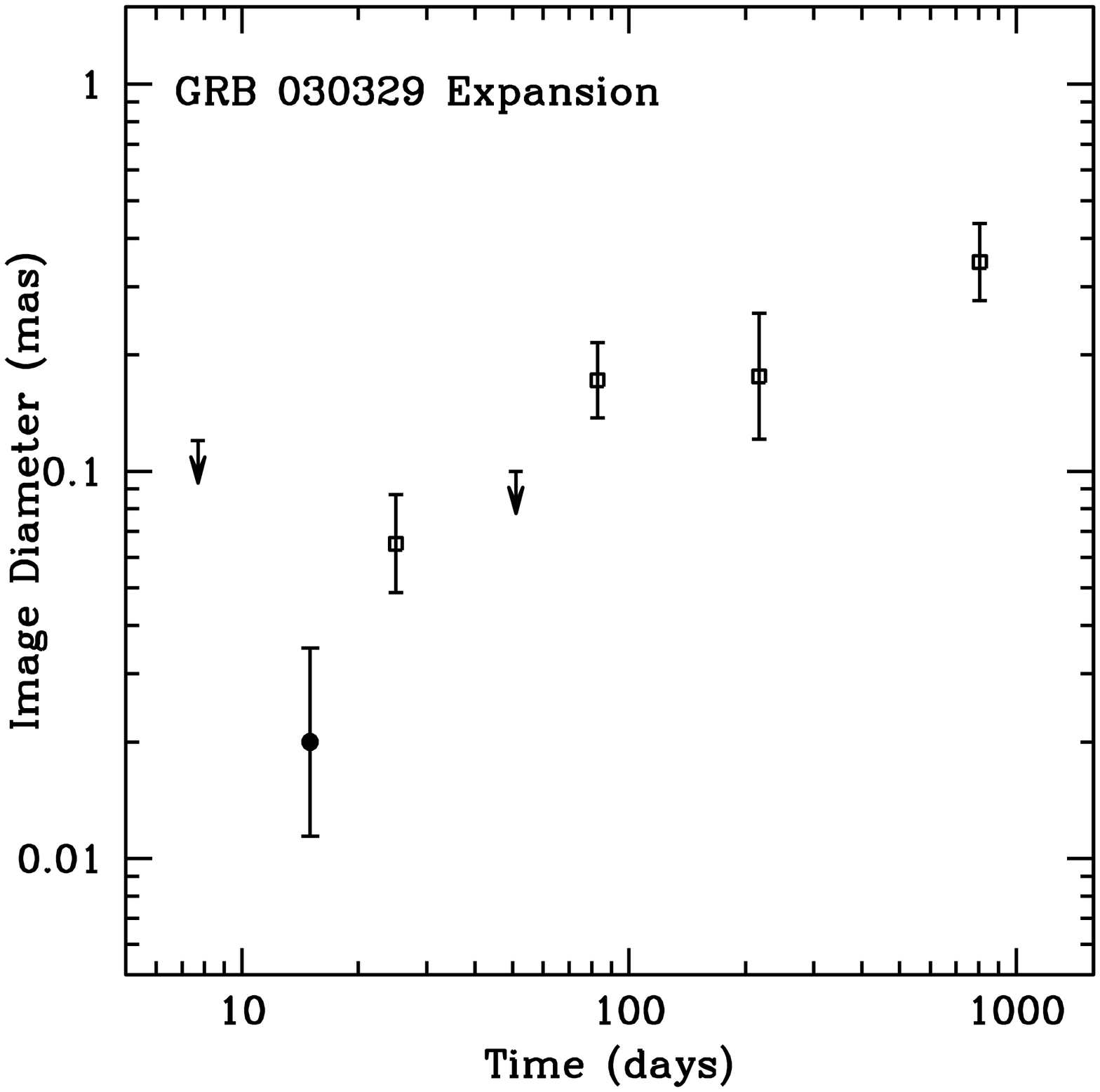}
\hspace{0.5cm}
\epsfxsize=6.8cm \epsfbox{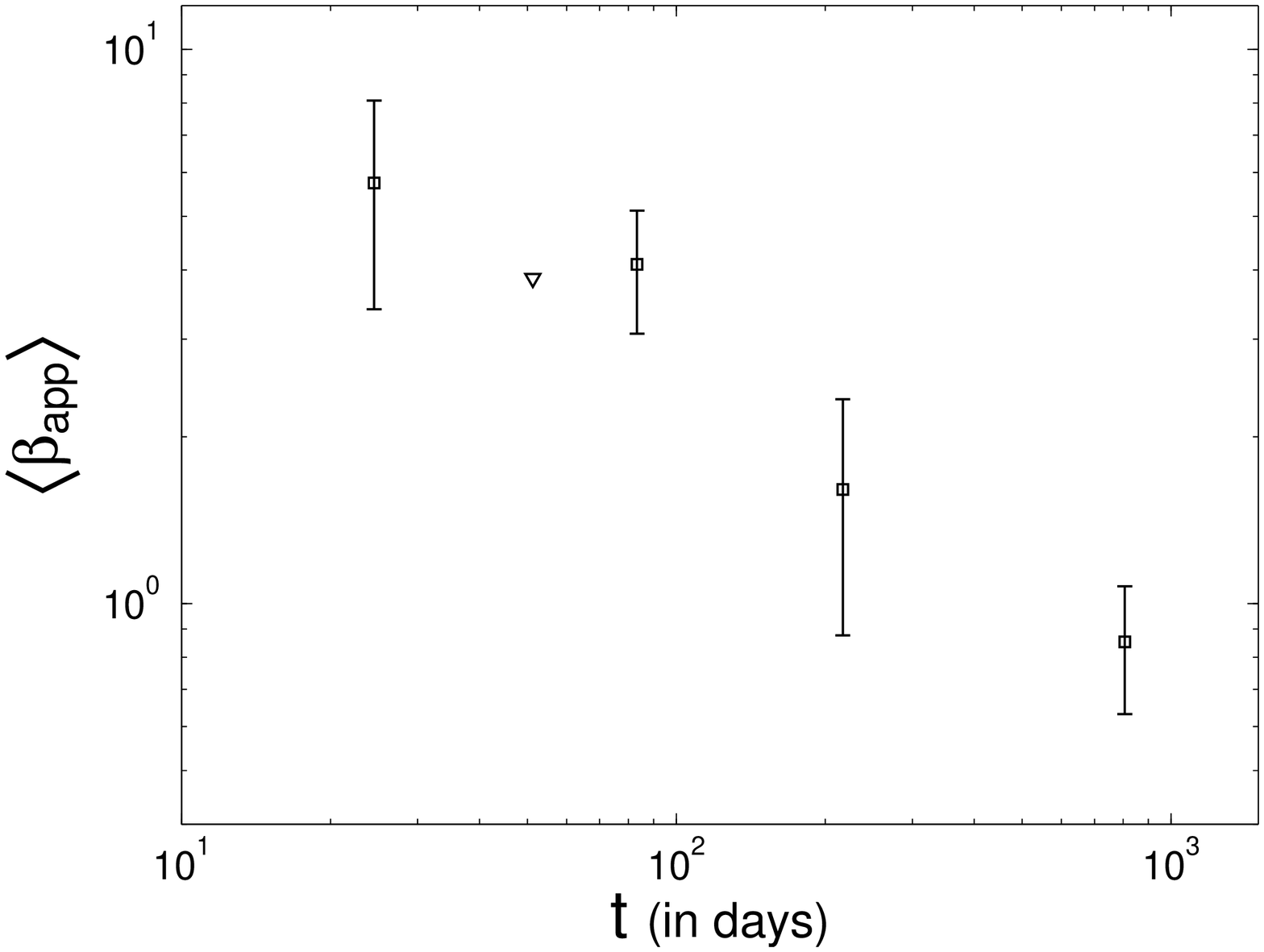}
%\end{center}
\caption{The temporal evolution of the radio afterglow image size 
of GRB\,030329 ({\it left panel}) and the implied average apparent
expansion velocity in units of $c$ ({\it right panel}). (Taken from
Pihlstr\"om et~al.\ 2007).}
\label{GRB030329_size}
\end{figure*}

A more direct measurement of the image size, as well as its temporal
evolution, can be obtained through very long base-line
interferometric techniques at radio wavelengths using, e.g., the Very
Long Baseline Array (VLBA). This has been possible for only one radio
afterglow so far - GRB\,030329 at a redshift of $z = 0.1685$ (Taylor et
al. 2004, 2005, Pihlstr\"om et~al.\ 2007), since it requires a bright,
relatively nearby event ($z \lesssim 0.2$). Nevertheless, the measured
source size and its temporal evolution imply a relativistic and
decelerating apparent expansion velocity (Fig.~\ref{GRB030329_size}),
with an average value of $\sim 6c$ over the first $25\,$days, and
transition to a sub-relativistic expansion after about one year.

\section{Aberration of light and the Doppler effect}
\label{beaming}

\begin{figure*}
\begin{center}
\epsfxsize=10.0cm \epsfbox{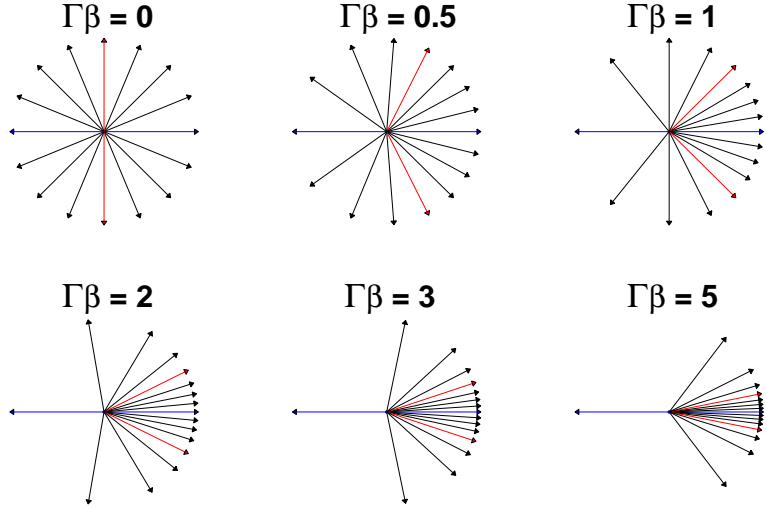}
\end{center}
\caption{Aberration of light: the arrows show the directions 
of photons in the lab frame for a point source that emits
isotropically in its own rest frame and moves to the right at
different values of the four-velocity $\Gamma\beta$. For $\Gamma \gg
1$, half of the photons (and 3/4 of the radiated energy) are within an
angle of $1/\Gamma$ around the source's direction of motion (between
the grey arrows, which correspond to $\theta' = 90^\circ$).}
\label{rel_beaming}
\end{figure*}

In relativistic sources such as GRBs, aberration of light (also known
as relativistic beaming) and the Doppler effect play an important
role.  These two effects refer to the change in direction and
frequency, respectively, of electromagnetic waves (or particles --
photons) between different frames of reference -- in our case between
the rest frame of the emitting fluid (that is referred to as the
comoving frame), and the lab frame (the rest frame of the central
source in which the external medium is at rest and the emitting jet
material is moving relativistically). These effects can be easily derived
from the Lorentz transformation of the 4-vectors of an
electromagnetic wave  $k^\mu = (\omega/c,\vec{k})$ or from the photon
energy-momentum $P^\mu = \hbar k^\mu = E(1/c,\hat{k})$. Some more
intuition may be gained by the following derivations using the Lorentz
transformation: $t = \Gamma(t'+\beta x'/c)$, $x = \Gamma(x'+\beta c
t')$, $y = y'$, $z = z'$. This implies 
\begin{equation}
v_x \equiv \frac{dx}{dt} = \frac{dx'+\beta c dt'}{dt' + \beta dx'/c}
= \frac{v'_x + \beta c}{1 + \beta v'_x/c}\ .
\end{equation}
For a photon, $v_x/c = \cos\theta \equiv \mu$ and $v'_x/c =
\cos\theta' \equiv \mu'$ are the cosines of the angle between its 
direction of motion and the $x$-direction (defined as the direction of
motion of the primed frame relative to the un-primed frame) in the two
rest frames, respectively. Therefore,
\begin{equation}\label{aberation_of_light}
\mu = \frac{\mu'+\beta}{1+\beta\mu'}\ ,\quad \mu' = \frac{\mu-\beta}{1-\beta\mu}\ ,
\end{equation}
where the second equality is derived from symmetry (interchanging
primed and un-primed quantities and replacing $\beta$ by
$-\beta$). This is the formula for aberration of light (or
relativistic beaming). It is demonstrated in Fig.~\ref{rel_beaming}
for a point source emitting isotropically in its own rest frame. For a
relativistic source moving with a Lorentz factor $\Gamma =
(1-\beta^2)^{-1/2} \gg 1$ (in the lab frame), half of the photons and
most ($3/4$) of the emitted energy are within an angle of $1/\Gamma$
around its direction of motion.

The formula for the Doppler factor, $\mathcal{D}\equiv \nu/\nu'$, can
be derived noticing that the phase of an electromagnetic wave must be
Lorentz invariant, since if the electric and magnetic fields vanish in
one frame then they must vanish in all frames. In particular, over one
period of oscillation of the electromagnetic field, $P = 1/\nu =
\lambda/c$, a photon travels a distance equal to its wavelength
$\lambda$, and similarly $P' = 1/\nu' = \lambda'/c$, so that
$\mathcal{D} = \lambda'/\lambda$.  Note that because of this simple
consideration $\lambda$ must be equal to the difference in the path
length to the observer (measured simultaneously in the lab frame)
between two hypothetical photons emitted with a single period time
difference (i.e., $c$ times the difference in the arrival time to the
observer of such hypothetical photons). Over a single period, a
time $P' =\lambda'/c$ elapses in the comoving frame, which due to time
dilation corresponds to a time $\Delta t_{\rm lab} = \Gamma\lambda'/c$
in the lab frame over which the source travels a distance of $l =
\beta c\Delta t_{\rm lab} = \Gamma\beta\lambda'$. During the same time the 
first hypothetical photon has traveled a distance of $c\Delta t_{\rm
lab} = \Gamma\lambda'$, which implies, from simple geometrical
arguments, that $\lambda = \Gamma\lambda' - l\mu =
\lambda'\Gamma(1-\beta\mu)$ and therefore $\mathcal{D} = \lambda'/\lambda 
= 1/\Gamma(1-\beta\mu)$.

Due to symmetry considerations similar to those mentioned above,
$\mathcal{D}' \equiv \nu'/\nu = 1/\Gamma(1+\beta\mu')$ and therefore
$\mathcal{D} = 1/\mathcal{D}' = \Gamma(1+\beta\mu')$. In particular,
$1/\Gamma(1-\beta\mu) = \Gamma(1+\beta\mu')$, which is equivalent to
eq.~\ref{aberation_of_light}. Expressing $\mathcal{D}$ in the comoving
frame makes it easier to calculate various quantities for an isotropic
emitter in this frame, which is broadly expected in most cases of
astrophysical interest. For example, it is easy to see that the
average Doppler factor for such an isotropic emitter is simply
$\Gamma$, while a fraction $(2+\beta)/4$ of the energy in the lab
frame is carried by photons with $\theta' < 90^\circ$ that correspond
to $\cos\theta >\beta$ (or $\theta < 1/\Gamma$ for $\Gamma \gg 1$).

Finally, we note that due to symmetry (no preferred direction in the
plane normal to $\vec{\beta}=\vec{v}/c$), the azimuthal angle of the photon
direction of motion is equal in the two frames, $\phi =\phi'$, and
thus the differential solid angles $d\Omega =d\phi d\mu$ and
$d\Omega' = d\phi'd\mu'$ are related by $d\Omega'/d\Omega = d\mu'/d\mu
= \mathcal{D}^2$, using eq.~\ref{aberation_of_light}.

\section{Evidence for jets in gamma-ray bursts}
\label{evidence_for_jets}

In contrast to Active Galactic Nuclei (AGN) and Galactic micro-quasars, in
many of which the relativistic jets are well resolved and their
structure can be studied directly, the images of GRBs are usually
unresolved, or at best the image of their late time radio afterglow is
only marginally resolved (i.e., the image size is still somewhat
smaller than the instrumental beam size, and its size and shape can be
only rather crudely estimated). Therefore, the evidence for jets
(i.e., highly collimated outflows) in GRBs is mainly indirect. The main
arguments, or lines of evidence, in favor of jets in GRBs are as
follows.

First, other known sources of relativistic outflows (such as
micro-quasars and AGN) are collimated into narrow jets, so it is
natural to expect a similar behavior for the relativistic outflow in
GRBs, if indeed the underlying process for launching jets are similar
(e.g., accretion onto a black hole).

Second, the very high values deduced for the energy output in gamma
rays assuming isotropic emission, $E_{\rm\gamma,iso}$, that are
inferred for GRBs with known redshifts, approach and in some cases
even exceed $M_\odot c^2$ (for GRB\,080916C $E_{\rm\gamma,iso} \approx
4.9M_\odot c^2$; Abdo et~al.\ 2009a). Such extreme energies in an
ultra-relativistic outflow are hard to envision in models involving
stellar mass progenitors. If the outflow is collimated into a narrow
jet (or bipolar jets) that occupies a small fraction $f_b \ll 1$ of
the total solid angle, then the strong relativistic beaming due to the
very high initial Lorentz factor ($\Gamma_0 \gtrsim 100$) will cause
the emitted gamma rays to be similarly collimated. This reduces the
true energy output in gamma rays by a factor of $f_b^{-1}$ to
$E_\gamma = f_b E_{\rm\gamma,iso}$, thus significantly reducing the
energy requirements (Rhoads 1997, Halpern et~al.\ 1999, Sari et~al.\
1999). In addition, there is good (spectroscopic) evidence that at
least some GRBs of the long-soft class (Kouveliotou et~al.\ 1993) are
coincident (to within a few days) with a core collapse supernova
belonging to the Type Ic category (Stanek et~al.\ 2003, Hjorth et~al.\
2003). In such cases the {\it average} Lorentz factor must be
$\langle\Gamma\rangle
\lesssim 2$ for a spherical explosion, since the accreted mass is not
expected to significantly exceed the ejected mass, and only a fraction
of the rest energy of the former can provide the kinetic energy for
the latter. Therefore, only a minute fraction of the ejected mass can
reach $\Gamma \gtrsim 100$ that is required in order to power the GRB.
Hydrodynamic analysis of spherical blastwaves (Tan et~al.\ 2001, Perna
\& Vietri 2002) shows that material with $\Gamma \gtrsim 100$ would
carry only a small fraction of the total energy, insufficient to
account for the high observed values of $E_{\rm\gamma,iso}$. This is
because the energy in the explosion is deposited into the bulk of the
ejecta, and only a decreasing fraction of the total energy is
transfered to a smaller fraction of the ejected mass that is
accelerated to subsequently higher velocities as the supernova shock
propagates into the steep density gradient at the outer edge of the
progenitor star. For a jet the ejected mass can be much smaller than
the accreted mass so that $\langle\Gamma\rangle
\gg 1$ is possible, and a good fraction of the total energy in the
outflow may be deposited in material with $\Gamma
\gtrsim 100$ that can power the GRB. Moreover, the energy requirements
for a jet are much less severe, since the same observed value of
$E_{\rm\gamma,iso}$ implies a much smaller beaming corrected radiated
energy $E_\gamma$, so that typically only a small fraction ($\lesssim
10^{-3}-10^{-2}$) of the total available energy is required to power
the jet.

Finally, a somewhat more direct line of evidence in favor of a
narrowly collimated outflow comes from achromatic breaks seen in the
afterglow light curves of many GRBs (Fruchter et~al.\ 1999, Kulkarni
et al.\ 1999, Harrison et~al.\ 1999, Halpern et~al.\ 2000, Price et
al.\ 2001, Sagar et~al.\ 2001).\footnote{Somewhat earlier, GRB\,980519
showed a steep afterglow flux decay rate ($\sim t^{-2}$), which was
interpreted as being due an earlier (unseen) jet break (Halpern et
al. 1999)} In fact, such a ``jet break'' in the afterglow light curve
was predicted before it was observed (Rhoads 1997, 1999, Sari et~al.\
1999). The exact cause of the jet break is discussed in detail in
Sect.~\ref{jet_break}. A brief explanation follows. Due to
relativistic beaming (and quasi-radial velocities of the emitting
material) most of the observed emission comes from a visible region of
angle $\sim 1/\Gamma$ around our line of sight. As the jet decelerates
by sweeping-up the external medium, the angular size of the visible
region increases, until the edge of the jet becomes visible when
$\Gamma$ drops below the inverse of its half-opening angle, and from
this point onward the observer ``misses'' the flux from outside the
edge of the jet, which would have been present if the flow were
spherical, resulting in a faster flux decay (i.e., a jet break).

For a bipolar jet that is uniform within its initial half-opening
angle, $\theta_0$, the collimation or beaming factor is given by $f_b
= 1-\cos\theta_0 \approx \theta_0^2/2$, and values of $f_b\sim
10^{-3}-10^{-1}$ (corresponding to $\theta_0$ values ranging between a
few degrees and a few tens of degrees) have been derived in the
pre-{\it Swift} era from such jet breaks (e.g., Frail et~al.\ 2001).
However, over the first years of {\it Swift} operations, only a
handful of convincing jet-break candidates were identified (Blustin et
al. 2006, Stanek et~al.\ 2007, Willingale et~al.\ 2007, Kocevski \&
Butler 2008), leading to concerns about the viability of this picture.
Since then, however, deep optical and X-ray observations have revealed
evidence for jet breaks in several additional {\it Swift} afterglows
(Dai et~al.\ 2008; Liang et~al.\ 2008; Racusin et~al.\ 2009; Burrows
et~al.\ 2009; Tanvir et~al.\ 2010). A significant fraction of these
aftreglows were monitored to late times, suggesting that jet breaks
for typical {\it Swift} GRBs may be occurring at late times and faint
flux levels that are beyond the limit of the standard ground- and
space-based campaigns. This, in turn, might be attributed to {\it
Swift}'s higher sensitivity compared to previous instruments, which
results in the detection of fainter GRBs that (on average) correspond
to wider jets.

\section{The jet structure}
\label{jet_structure}

The lack of well resolved GRB images makes it very difficult for the
jet structure to be deduced. The initial Lorentz factor during the
prompt gamma-ray emission is very high, $\Gamma_0 \gtrsim 100$, and
therefore we observe emission mainly from very small angles, $\theta
\lesssim \Gamma_0^{-1} \lesssim 10^{-2}\,$rad, relative to our line of
sight. This is a result of relativistic beaming (i.e., aberration of
light;
see Sect.~\ref{beaming}).  For this reason, the prompt
gamma-ray emission probes only a very small region of solid angle $\sim
\pi\Gamma_0^{-2}$, or a fraction $\sim \Gamma_0^{-2}/4 \sim 10^{-7} -
10^{-4.5}$ of the total solid angle. Thus, the prompt emission
provides no information about the ejecta that are moving in other
directions (i.e., at angles $\gg\Gamma_0^{-1}$ from our direction): the
outflow could be spherical, or concentrated in a conical jet of
half-opening angle $\theta_0 > \Gamma_0^{-1}$ (provided that our line
of sight is inside the cone, at an angle of $\gtrsim \Gamma_0^{-1}$
from its edge). 

During the afterglow, however, the Lorentz factor $\Gamma$ of the
emitting material decreases with time, since the afterglow shock
decelerates as it sweeps up the external medium. This causes the
beaming (or light aberration) effects to gradually become less severe,
as $\Gamma$ becomes more moderate, so that we observe afterglow
emission from a wider range of angles (of $\lesssim \Gamma^{-1}$ from
our line of sight, and not just from material moving almost directly
towards us, at angles $\lesssim \Gamma_0^{-1}$, as in the prompt
emission). This enables us to probe the jet structure over
increasingly larger angular scales.

For a jet with axial symmetry, its structure~\footnote{We refer here
to the angular structure, and ignore the radial structure of the
outflow, which becomes unimportant once most of the energy is
transfered to the shocked external medium.} can be described by the
distribution of its total energy (excluding rest energy) content per
solid angle, $\mathcal{E}$, and initial Lorentz factor $\Gamma_0$, as
a function of the jet's polar angle $\theta$.  We note that
$\Gamma_0(\theta)$ affects mainly the prompt gamma-ray emission and
early afterglow, since it is largely ``forgotten'' after the local
deceleration time $t_{\rm dec}(\theta) 
\sim R_{\rm dec}(\theta)/2c\Gamma_0^2(\theta)$, while 
$\mathcal{E}(\theta)$ affects also the late time afterglow
emission. Here $R_{\rm dec}(\theta)$ is the local deceleration radius
where most of the energy in the outflow at an angle $\theta$ is
transferred to the shocked external medium; for a power-law external
density, $\rho_{\rm ext}(r) = Ar^{-k}$, it is given by~\footnote{This
expression assumes a ``thin'' shell, as discussed in
Sect.~\ref{dynamics}.}  $R_{\rm dec}(\theta) \approx
[(3-k)\mathcal{E}(\theta)/Ac^2\Gamma_0^2(\theta)]^{1/(3-k)}$.  The
structure of GRB jets is very important for deducing their event rate
and total energy, as well as for requirements on the jet production
mechanisms.

The most popular model for the structure of GRB jets is the uniform
jet (UJ, or ``top hat'') model (Rhoads 1997, 1999, Panaitescu \&
M\'esz\'aros 1999, Sari et~al.\ 1999, Kumar \& Panaitescu
2000, Moderski et~al.\ 2000, Granot et~al.\ 2001, 2002,
Ramirez-Ruiz \& Madau 2004, Ramirez-Ruiz et~al.\ 2005), where
$\mathcal{E}$ and $\Gamma_0$ are uniform within some finite
half-opening angle, $\theta_{\rm j}$, and sharply drop outside of
$\theta_{\rm j}$. An alternative jet structure that is also rather
popular is the universal structured jet (USJ) model (M\'esz\'aros
et~al.\ 1998, Lipunov et~al.\ 2001, Rossi et~al.\ 2002, Zhang \&
M\'esz\'aros 2002), where $\mathcal{E}$ and $\Gamma_0$ vary smoothly
with $\theta$, as a power law outside of some narrow core angle,
typically with equal energy per decade in $\theta$, $\mathcal{E}
\propto \theta^{-2}$. In the UJ model the different values of the jet
break time, $t_{\rm j}$, in the afterglow light curve arise mainly due
to different $\theta_{\rm j}$ (and to a lesser extent due to different
ambient densities). In the USJ model, all GRB jets are assumed to be
identical, and the different values of $t_{\rm j}$ arise mainly due to
different viewing angles, $\theta_{\rm obs}$, from the jet axis. In
fact, the expression for $t_{\rm j}$ is similar to that for a uniform
jet with $\mathcal{E} \to
\mathcal{E}(\theta=\theta_{\rm obs})$ and $\theta_{\rm j} \to \theta_{\rm
obs}$.

An alternative jet structure that has been proposed in the literature
is one with a Gaussian angular profile (Zhang \& M\'esz\'aros 2002,
Kumar \& Granot 2003). It may be thought of as a more realistic
version of a uniform jet, where the edges are smooth rather than
sharp. A Gaussian jet, $\mathcal{E}(\theta) \propto
\exp(-\theta^2/2\theta_c^2)$, is approximately intermediate between
the UJ and USJ models. However, it is closer to the UJ model than to
the USJ model with $\mathcal{E} \propto \theta^{-2}$ in the sense that
the energy in the wings of a Gaussian jet is much smaller than in its
core, whereas for a USJ with $\mathcal{E}\propto\theta^{-2}$ wings
there is equal energy per decade (in $\theta$) in the wings, and
therefore the wings contain more energy than the core (by about an
order of magnitude).

Another jet structure that is gradually receiving more attention is a
two-component jet (Pedersen et~al.\ 1998, Frail et~al.\ 2000, Berger
et al.\ 2003, Huang et~al.\ 2004, Peng et~al.\ 2005, Racusin et al.\
2008) with a narrow uniform jet of initial Lorentz factor $\Gamma_0
\gtrsim 100$ surrounded by a wider uniform jet with $\Gamma_0 \sim 10
- 30$. Such a jet structure was predicted in the context of the cocoon
in the collapsar model (Ramirez-Ruiz et~al.\ 2002) and in the context
of a hydromagnetically driven neutron-rich jet (Vlahakis et~al.\
2003). This model has been invoked in order to account for sharp bumps
(i.e., fast rebrightening episodes) in the afterglow light curves of
GRB\,030329 (Berger et~al.\ 2003) and XRF\,030723 (Huang et al.\
2004), but detailed calculations show that it cannot produce very
sharp features in the light curve (Granot 2005). A different
motivation for such jet structure comes about from the energetics of
GRBs and X-ray flashes, which could help reduce the high efficiency
requirements from the prompt gamma-ray emission (Peng et~al.\
2005). Later {\it Swift} observations (e.g., Nousek et~al.\ 2006)
showed that while it can reproduce the early X-ray afterglow light
curves, and specifically the shallow decay phase, it does not
significantly help reduce the required gamma-ray efficiency (Granot
et~al.\ 2006). A different jet structure that has been suggested is
that of a jet with a cross section in the shape of a ``ring,''
sometimes referred to as a ``hollow cone'' (Levinson \& Eichler 1993,
2000, Eichler \& Levinson 2003, 2004, Lazzati \& Begelman 2005), which
is uniform within $\theta_c < \theta < \theta_c + \Delta\theta$ where
$\Delta\theta \ll \theta_c$.

Finally, it has been argued that there might be random variations on
small angular scales in the prompt GRB brightness and in the energy
per solid angle in the jet (and therefore also in the corresponding
afterglow brightness), around some uniform mean value -- the ``patchy
shell'' model (Kumar \& Piran 2000). This model predicts wide
variations in $E_{\rm\gamma,iso}$ between different lines of sight
relative to the same GRB jet, as well as fluctuations in the afterglow
light curve whose amplitude decreases with time as the visible region
(of angle $\lesssim \Gamma^{-1}$ around the line of sight) increases,
thus effectively averaging the emission over an increasing number of
bright and dim regions. The afterglow flux for different lines of
sight approaches the same value at late times, thus implying a
relatively narrow distribution of the afterglow (isotropic equivalent)
luminosity compared to $E_{\rm\gamma,iso}$, and no obvious correlation
with $E_{\rm\gamma,iso}$. While this seemed plausible at the time,
later {\it Swift} observations showed a clear correlation where GRBs
with large $E_{\rm\gamma,iso}$ tend to have a more luminous afterglow
emission (Nousek et~al.\ 2006, Gehrels et~al.\ 2008).  A more extreme
version of the ``patchy shell'' model is the ``mini-jets'' model
(Yamazaki et~al.\ 2004), where the regions of bright emission and high
energy per solid angle are considered as discrete mini-jets, and the
regions between them are assumed to have a negligible emission or
energy per solid-angle. The main difference is that in this model many
lines of sight fall between the mini-jets and the early emission is
dominated by the one or more mini-jets closest to the line of sight,
whose beaming cone initially does not encompass the line of sight. The
original motivation for this model was unifying short GRBs with long
GRBs and XRFs, where long GRBs, short GRBs and XRFs correspond to
lines of sight with many, one and no mini-jets, respectively. However,
the later discovery that the host galaxies of short GRBs are a
different population from those of long GRBs (Gehrels et~al.\ 2005,
see also Fong et~al.\ 2010) rules out the inclusion of short GRBs in
such a unification scheme.\\

\noindent
{\bf Constraining the jet structure:} efforts have been made to
constrain the jet structure through statistical studies, linear
polarization, afterglow light curves, off-axis viewing and orphan
afterglows. Statistical studies have focused on the $\log N - \log S$
distribution where $N$ is the number of GRBs above some threshold peak
photon flux $S$ (e.g., Firmani et~al.\ 2004, Guetta et~al.\
2005), as well as on $dN/d\theta$ (Perna et~al.\ 2003) or
$dN/dzd\theta$ (Nakar et~al.\ 2004), where $z$ is the
redshift and $\theta$ is $\theta_{\rm j}$ for the UJ model or $\theta_{\rm
obs}$ for the USJ model, and is determined from the jet break time in
the afterglow light curve. These studies were inconclusive, though
they showed some preference for the UJ model over the USJ model. The
same holds for studies involving the shape of afterglow light curves
(Granot \& Kumar 2003), but nevertheless some jets structures could be
ruled out using this method (Granot 2005). Studies based on afterglow
polarization evolution need to deconvolve the effects of the jet
structure from those of the magnetic field configuration in the
post-shock emitting region, which make it difficult to draw strong
conclusions about the jet structure without making similarly strong
assumptions about the structure of the magnetic field itself.

Orphan afterglows are events in which the late time afterglow is
detected while the prompt gamma-ray emission is not. Unfortunately, no
such events have been clearly observed so far, and the upper limits on
the rates of orphan afterglows are still not very constraining for
most of the different jet structures. Nevertheless, upper limits on
the fraction of SN Ib/c that show late time radio emission of possible
afterglow origin (Soderberg et~al.\ 2006) already argue quite strongly
against the presence of energetic, narrow jets (of half-opening angle
$\theta_0 \sim 1/\Gamma_0 \sim 10^{-2.5}$) which would imply that a
large fraction SNe Ib/c should harbor GRB jets (most of which point
away from us and can be detected only at late times in the radio, when
the jets become sub-relativistic). Such extremely narrow jets would
also imply a very low true energy, which would be inconsistent with
the lower limits on the kinetic energy inferred from late time radio
afterglow observations (Berger, Kulkarni \& Frail 2004, Frail et
al. 2005). On the other hand, two Type Ic supernovae have been
recently observed to harbor relativistic jets (SN~2007gr -- Paragi et
al. 2010, and SN 2009bb -- Soderberg et~al.\ 2010), albeit with an
energy either much (SN~2007gr) or slightly (SN~2009bb) lower than is
needed to produce a {\it typical} bright GRB. These two supernovae
amount to about $\sim 1\%$ of relatively nearby Type Ib/c supernovae,
for which a search for such jets has been performed at radio
wavelengths.

%%%%%%%%%%%%%%%%%%%%%%%%%%%%%%%%%%%%%%CHANGES STOP%%%%%%%%%%%%%%%%
\section{Dynamics of gamma-ray burst outflows}
\label{dynamics}

Here we provide a brief summary of the key aspects for the dynamics of
the interaction between the GRB outflow and the surrounding medium,
and provide a simple intuitive explanation for some of the most
important results. We begin with the relatively simpler case of a
spherical outflow, and consider a uniform unmagnetized shell of ejecta
of initial width $\Delta_0$ and Lorentz factor $\Gamma_0$ in the lab
frame, propagating into an external density $\rho_1(R)$. A forward
shock is driven into the ambient medium, while the ejecta are
decelerated by a reverse shock.  Thus four regions exist: (1) unshocked
external medium, (2) shocked external medium, (3) shocked ejecta, and
(4) unshocked (freely expanding) ejecta. All the velocities are
measured relative to region 1, while the pressure $p$ and rest mass
density $\rho$ (or number density $n$) are measured in the fluid rest
frame. A subscript $i$ between 1 and 4 refers to region $i$, while a
subscript $ij$ refers to the relative velocity of regions $i$ and $j$,
so that $\Gamma_{i1} = \Gamma_{1i} = \Gamma_{i}$. Given $\Gamma_4 =
\Gamma_0$, $\Gamma_1 = 1$, $\rho_4$, $\rho_1$, and assuming the shell 
of ejecta and external medium are both cold ($p_i \ll \rho_i c^2$ for
$i = 1,\,4$), there are 8 unknown hydrodynamic quantities ($\rho_2$,
$p_2$, $\Gamma_2$, $\rho_3$, $p_3$, $\Gamma_3$, and the Lorentz
factors of the forward and reverse shock fronts) that can be found
from the conditions across the contact discontinuity separating
regions 2 and 3 ($p_2 = p_3$ and $\Gamma_2 = \Gamma_3 \equiv \Gamma$)
as well as the shock jump conditions (continuity of the energy,
momentum, and particle fluxes) across the forward and reverse shocks
(between regions 1 \& 2, and 4 \& 3, respectively). For simplicity,
this is treated in planar symmetry, and the spherical nature of the
flow enters only when the evolution of the flow with radius is
considered.

An approximate, order of magnitude, estimate of the dynamics at this
stage can be obtained by equating the ram pressure of the incoming
fluid from regions 4 and 1, as seen from the contact discontinuity
(the rest frame of regions 2 and 3): $\rho_1 u_{21}^2 \sim
\rho_4 u_{34}^2$, where $u = \Gamma\beta$ is the 4-velocity. 
For $\rho_4 = \rho_1$ we must have $u_{21} = u_{34}$ due to symmetry,
and for $\Gamma_{4} = \Gamma_0 \gg 1$ this implies that both the
forward and reverse shocks are relativistic, $u_{21} = u_{43} =
[(\Gamma_{4}-1)/2]^{1/2} \approx (\Gamma_{4}/2)^{1/2} \gg 1$.  As long
as the forward shock is relativistic, $1 \ll \Gamma_3 = \Gamma_2
\approx u_{21} \sim u_{34}(\rho_4/\rho_1)^{1/2}$; when the reverse
shock is relativistic then $1 \ll u_{43} \approx \Gamma_{43} \approx
\Gamma_{4}/2\Gamma_3$, so that $\Gamma_2 \approx
(\Gamma_0/2)^{1/2}(\rho_4/\rho_1)^{1/4}$ and $\Gamma_{34} \approx
(\Gamma_0/2)^{1/2}(\rho_1/\rho_4)^{1/4}$ (Sari \& Piran
1995). Therefore, the condition for the forward shock to be
relativistic is $\Gamma_0^2 \gg \rho_1/\rho_4$, which is typically
always satisfied (until very late times when the flow becomes
Newtonian), so that $u_{21} \approx \Gamma_2$ and $u_{43} \sim
\Gamma_2(\rho_1/\rho_4)^{1/2}$. The condition for the reverse shock to
be relativistic is $\Gamma_0^2 \gg \rho_4/\rho_1$.  The revere shock
is Newtonian in the opposite limit, $\Gamma_0^2 \ll \rho_4/\rho_1$,
and, in this case, $\Gamma_2 \approx \Gamma_0$ so that $u_{43} \sim
\Gamma_0(\rho_1/\rho_4)^{1/2} \ll 1$.

Note that $u_{43} \sim \min(a^{1/4},a^{1/2})$ where $a =
\Gamma_0^2\rho_1/\rho_4 \sim A c^2 \Gamma_0^4 R^{2-k}\Delta/E \sim
\Gamma_0^4R^{2-k}\Delta/l^{3-k} $ for a power-law external density,
$\rho_1 = A R^{-k}$, where $l \sim (E/Ac^2)^{1/(3-k)}$ is the Sedov
length. For a narrow distribution of Lorentz factors $\Delta \approx
\Delta_0 = {\rm const}$, and therefore $a \propto R^{2-k}$, i.e., for 
$k < 2$ the reverse shock is initially Newtonian and strengthens with
radius.  If there is a reasonable spread in the Lorentz factor of the
outflow, $\Delta \Gamma_0 \sim \Gamma_0$, then the shell can spread,
$\Delta \sim \max(\Delta_0,R/\Gamma_0^2)$, where $\Delta \propto R$
and therefore $a \propto R^{3-k}$ at $R > R_{\rm s} \sim \Gamma_0^2\Delta_0$
so that then the reverse shock strengthens for $k < 3$.

If the reverse shock is relativistic by the time it finishes crossing
the shell, then most of the energy is transfered to the shocked
external medium within a single shell crossing. If there is only a
very small spread in $\Gamma_0$ then the reverse shock can still be
Newtonian when it finishes crossing the shell. In this case a large
number of Newtonian shocks and rarefaction waves may need to cross the
shell before most of the energy is transfered to the shocked external
medium (Sari 1997). However, if there is a reasonable spread in the
Lorentz factor of the outflow, $\Delta \Gamma_0 \sim \Gamma_0$, then
the shell starts to spread before the reverse shock finishes crossing
it, in such a way that by the time it crosses the shell the reverse
shock already becomes mildly relativistic, so that most of the energy
is transfered to the shocked external medium in a single shell
crossing time (Sari \& Piran 1995). The dividing line between these
two cases corresponds to $1 = a(R_{\rm s}) \sim
\Gamma_0^{2(4-k)}(\Delta_0/l)^{3-k}$, where $a(R_{\rm s}) > 1$ implies a
relativistic reverse shock or a ``thick'' shell and $a(R_{\rm s}) < 1$
implies a Newtonian reverse shock (without spreading) or a ``thin''
shell.

Most of the energy is transferred to the shocked external medium at a
radius $R_{\rm dec} \sim l\Gamma_{\rm dec}^{-2/(3-k)}$ where
$\Gamma_{\rm dec} = \Gamma(R_{\rm dec}) \sim \min[\Gamma_0,\Gamma_{\rm
cr}]$ and
\begin{equation}
\Gamma_{\rm cr} 
=\left(\frac{l}{\Delta_0}\right)^{(3-k)/2(4-k)}
=\left\{\matrix{
280\,\zeta^{3/8}E_{53}^{1/8}n_0^{-1/8}T_{50}^{-3/8}
& (k=0)\ , \cr & \cr
70\zeta^{1/4}E_{53}^{1/4}A_*^{-1/4}T_{50}^{-1/4}
& (k=2)\ , }\right.
\end{equation}
where $\zeta=(1+z)/2$, $T_{\rm GRB}= (1+z)\Delta_0/c=50T_{50}\,$s is
the duration of the GRB, $E = 10^{53}E_{53}\,$erg is the (isotropic
equivalent) energy of the flow, and $n=n_0\,{\rm cm^{-3}} = A/m_p$ for
$k = 0$, while $A_*=A/(5\times 10^{11}\,{\rm gr\, cm^{-1}})$ for $k =
2$.  For $\Gamma_0 > \Gamma_{\rm cr}$ we have a thick shell or
relativistic reverse shock, and the observed deceleration time is
similar to the duration of the GRB, $t_{\rm dec} \sim R_{\rm
dec}/2c\Gamma_{\rm dec}^2 \sim (1+z)\Delta_0/c \sim T_{\rm GRB}$. For
$\Gamma_0 < \Gamma_{\rm cr}$ we have a Newtonian (or at most mildly
relativistic) reverse shock or a thin shell, and in this case $t_{\rm
dec} \sim (l/c)\Gamma_0^{-2(4-k)/(3-k)} > T_{\rm GRB}$ (where $l/c
\sim t_{\rm NR}$ is the non-relativistic transition time) and is given
by
\begin{equation}
t_{\rm dec} = (1+z)\frac{R_{\rm dec}}{2c\Gamma_0^2} = \left\{\matrix{
18\,\zeta E_{53}^{1/3}n_0^{-1/3}(\Gamma_0/10^{2.5})^{-8/3}\,{\rm s}
& (k=0)\ , \cr & \cr
5.9\zeta E_{53} A_*^{-1}(\Gamma_0/100)^{-4}\,{\rm s}
& (k=2)\ . }\right.
\end{equation}

Once most of the energy in the GRB outflow is transfered to the
shocked external medium, the flow becomes self-similar, and is
described by the Blandford \& McKee (1976) solution. The scaling of
the basic quantities with radius during this stage may be understood
as follows. In this stage most of the energy is in the shocked
external medium, and therefore the dynamics may be found by energy
conservation within this region. In the rest frame of the the shocked
external medium (region 2), the cold upstream external medium
approaches at a Lorentz factor $\Gamma_{12} = \Gamma_2$, which for
simplicity we denote here by $\Gamma$, and the velocities of the
particles are randomized at the shock font, such that the ordered bulk
motion in the upstream region is converted to random motion of the
particles in the downstream region (2), with the same average Lorentz
factor. Therefore, in the rest frame of the shocked downstream medium,
the average energy per particle (including its rest energy) is
$\Gamma$ times its rest energy. In the lab frame the energy is larger
by a factor of $\Gamma$, implying that the total kinetic energy in the
lab frame is $E = (\Gamma^2-1) M(R) c^2 = u^2 M(R) c^2$, where $M(R)$
is the total swept-up rest mass up to radius $R$, $u = \Gamma\beta$ is
the 4-velocity, and we have deducted the rest energy in order to
obtain the kinetic energy (only the latter is conserved, since new
rest-mass is added to the shocked region as the external medium is
swept-up by the forward shock).

For an external density that varies as a power law with radius,
$\rho_1 \propto R^{-k}$, we have $E \propto u^2 R^{3-k}$ and therefore
conservation of energy in the lab frame implies that $u \propto
R^{(k-3)/2}$. For a relativistic flow $u \approx \Gamma$ and $\Gamma
\propto R^{(k-3)/2}$.  Eventually the flow becomes Newtonian and
approaches the Sedov-Taylor solution. In this regime $u \approx \beta$
and therefore $\beta \propto R^{(k-3)/2}$. Both scalings given above
apply to the adiabatic case where energy losses (e.g., due to
radiation) or gains (e.g., due to late time energy injection from the
central source) can be neglected. For simplicity we consider here only
$k < 3$ for which $u$ decreases with $R$. The pressure in the shocked
region scales as $p \sim \rho_1 c^2 u^2 \propto R^{-3}$, while the
particle rest mass density (in the comoving frame) scales as $\rho_2
\sim \Gamma \rho_1$, i.e.,  as $R^{-(3+k)/2}$ in the relativistic case
and as $R^{-k}$ for the Newtonian case. A spherical flow becomes
non-relativistic once $\Gamma^2 \approx E/M(R)c^2 \sim 1$, at
\begin{equation}
t_{\rm NR} \sim \frac{l}{c} = \left[\frac{(3-k)E}{4\pi Ac^{5-k}}\right]^{1/(3-k)}
=\left\{\matrix{
5.3\,\zeta E_{53}^{1/3}n_0^{-1/3}\,{\rm yr}
& (k=0)\ , \cr & \cr
37\zeta E_{53} A_*^{-1}\,{\rm yr}
& (k=2)\ . }\right.
\end{equation}

Variations on the relatively simple dynamics described above after the
deceleration epoch involve relaxing either the assumption of a
constant energy or of spherical symmetry. The energy can decrease due
to radiative losses or increase due to energy injection, either from a
relativistic wind caused by prolonged activity of the central source,
or slow material that was ejected promptly from the source but
catches up with the shocked external medium as the latter decelerates
to a somewhat smaller Lorentz factor (e.g., Blandford \& McKee 1976,
Cohen \& Piran 1999, Sari \& M\'esz\'aros 2000, Ramirez-Ruiz et~al.\ 
2001). Here we discuss deviation from spherical symmetry, for
which the dynamics are more complicated and less certain. For
simplicity, we shall consider an axisymmetric double-sided jet that is
uniform within an initial half-opening angle of $\theta_0 \ll 1$
around its symmetry axis with sharp edges.

If a signal (e.g., a sound or rarefaction wave) can travel laterally
within the jet at a speed of $\beta_{\rm s} c$ in its comoving frame, then
it traverses an angle of $d\theta \approx dR_\perp/R = \beta_{\rm s} c dt'/R
\approx \beta_{\rm s} dR/R\Gamma$, so that even for $\beta_{\rm s} \sim 1$ (which
may indeed be expected) information traverses an angle of $\sim
1/\Gamma$ on the dynamical (or radius doubling) time. Therefore, the
center of the jet knows about its edge only when $\Gamma$ drops to
$\sim 1/\theta_0$; at earlier times it behaves as if it were part
of a spherical flow with the same external density profile and
isotropic equivalent energy $E_{\rm iso}$. Using the spherical
adiabatic scaling $\Gamma \propto R^{(k-3)/2}$ a signal starting at
$R_0$ traverses $\Delta\theta \approx 2\beta_{\rm s}/(3-k)\Gamma(R)$ by a
radius $R \gg R_0$ for $k < 3$. Therefore, when $\Delta\theta =
\theta_0$ corresponding to $\Gamma\theta_0 \approx 2\beta_{\rm s}/(3-k)$,
the center of the jet comes into causal contact with its edge, and the
jet can in principle start to spread sideways rapidly.

Early analytic works (Rhoads 1997, 1999, Sari et~al.\ 1999)
assumed that the jet half-opening angle grows as $\theta_{\rm j} = \theta_0
+ \Delta \theta \sim \theta_0 + \beta_{\rm s}/\Gamma$ and indeed quickly
expands laterally once $\Gamma$ drops below $\sim \beta_{\rm s}/\theta_0
\sim 1/\theta_0$, around the same time as the edges of the jet become 
visible~\footnote{Since the visible region is within an angle of $\sim
1/\Gamma$ around the line of sight due to relativistic beaming and
light travel effects.}. As a result, a steepening in the afterglow
light curve is produced, which is known as a jet break and was detected
soon after it was predicted. The jet break occurs at a radius $R_{\rm j}
\sim R_{\rm NR}(E) \sim \theta_0^{2/(3-k)}R_{\rm NR}(E_{\rm iso})$, 
and the simple analytic models (e.g., Rhoads 1999) suggest that at $R >
R_{\rm j}$, $\Gamma \sim \theta_0^{-1}e^{1-R/R_{\rm j}}$ and $\theta_{\rm j} \sim
\theta_0(R_{\rm j}/R)e^{1-R/R_{\rm j}}$ until $\Gamma \sim 1$ at $R \sim
R_{\rm j}(1-\ln\theta_0)$. Different analytic models vary in their exact
assumptions and results (e.g., Piran 2000, Granot 2007), but the
exponential decrease in $\Gamma$ and increase in $\theta_{\rm j}$ with
radius are a generic prediction. In this picture the jet becomes
sub-relativistic at close to $R_{\rm NR}(E)$ (to within a logarithmic
factor), and therefore the transition to the late time asymptotic
spherical Newtonian Sedov-Taylor solution is relatively prompt and
smooth in this scenario.

However, numerical studies (Granot et~al.\ 2001, Kumar \& Granot 2003,
Cannizzo et~al.\ 2004, Zhang \& MacFadyen 2009), and in particular full
special relativistic two dimensional hydrodynamic simulations, which
are the best and most reliable calculations to date, show that the
lateral expansion of the jet is very modest as long as it remains
relativistic. The big difference compared to the results of simple
analytic models may be attributed to the over-simplified assumptions,
such as the jet being perfectly uniform with sharp
edges within some finite half-opening angle with a purely radial
velocity, which are shown to be invalid in the numerical
simulations. Because of the lack of significant sideways expansion
during the relativistic stage, the gross properties of the jet may be
approximated to zeroth order by a conical section of half opening
angle $\sim \theta_0$ out of a spherical flow. Therefore, the jet
becomes non-relativistic at a radius $\sim R_{\rm NR}(E_{\rm iso})
\sim \theta_0^{-2/(3-k)}R_{\rm NR}(E)$, i.e., near the Sedov length 
corresponding to its initial isotropic equivalent energy $E_{\rm
iso}$, which for a narrow jet ($\theta_0 \ll 1$) is significantly
larger than the Sedov length corresponding to its true energy $E
\approx E_{\rm iso}\theta_0^2/2$. This causes the transition into 
a spherical flow described by the Sedov-Taylor solution (with the true
energy $E$) to extend over a large range of times, with a rather
modest growth in the (maximal) radius during this transition period
(Granot et~al.\ 2005).

\section{The afterglow emission}
\label{AG_emission}

The dominant emission mechanism during the afterglow stage is believed
to be synchrotron radiation, produced as relativistic electrons
accelerated by the afterglow shock gyrate in the magnetic fields
within the shocked external medium. A synchrotron origin of the
afterglow emission is supported by the detection of linear
polarization at the level of $\sim 1\%-3\%$ in several optical or NIR
afterglows (see Sect.~\ref{pol}), and by the shape of the broad band
spectrum, which consists of several power-law segments that smoothly
join at some typical break frequencies (Galama et~al.\ 1998). 
Synchrotron self-Compton (SSC)
-- the inverse-Compton scattering of the synchrotron photons to (much)
higher energies by the same population of relativistic electrons that
emits the synchrotron photons -- can sometimes dominate the afterglow
flux in the X-rays (Sari \& Esin 2001, Harrison et~al.\ 2001), and may
affect the synchrotron emission by increasing the electron cooling.

\begin{figure*}
%\vspace{0.1cm}
%\hspace{0.10cm}
\begin{center}
\epsfxsize=10.0cm \epsfbox{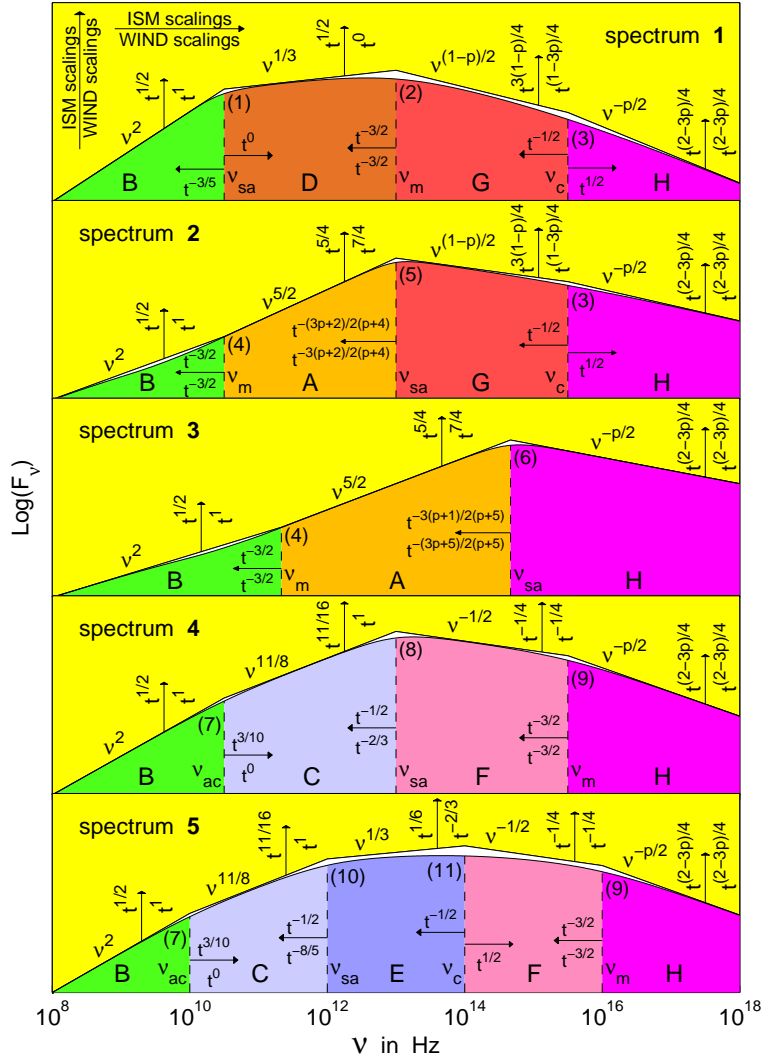}
\end{center}
\vspace{-0.3cm}
\caption{The afterglow synchrotron spectrum, calculated for the
Blandford \& McKee (1976) spherical self-similar solution, under
standard assumptions, using the accurate form of the synchrotron
spectral emissivity and integration over the emission from the whole
volume of shocked material behind the forward (afterglow) shock (for
details see Granot \& Sari 2002). The different panels show the five
possible broad band spectra of the afterglow synchrotron emission,
each corresponding to a different ordering of the spectral break
frequencies. Each spectrum consists of several power-law segments
(PLSs; each shown with a different color and labeled by a different
letter A--H) that smoothly join at the break frequencies (numbered
1--11). The broken power-law spectrum, which consists of the
asymptotic PLSs that abruptly join at the break frequencies (and is
widely used in the literature), is shown for comparison.  Most PLSs
appear in more than one of the five different broad band
spectra. Indicated next to the arrows are the temporal scaling of the
break frequencies and the flux density at the different PLSs, for a
uniform (ISM; $k=0$) and stellar wind (WIND; $k=2$) external density
profile.}
\label{fig:spectrum}
\end{figure*}

Since the physics of relativistic collisionless shocks (and in
particular how they amplify the magnetic field and accelerate
particles to a non-thermal relativistic energy distribution) are still
not well understood from first principles, simple assumptions are
usually made that conveniently parameterize our ignorance. The
electrons are usually assumed to be (practically instantaneously)
shock-accelerated into a power-law distribution of energies,
$dN/d\gamma_e \propto \gamma_e^{-p}$ for $\gamma_e > \gamma_m$, and
thereafter to cool both adiabatically and due to radiative losses. It
is further assumed that practically all of the electrons take part in
this acceleration process and form such a non-thermal (power-law)
distribution, leaving no thermal component (which is not at all clear
or justified; e.g., Eichler \& Waxman 2005). The relativistic electrons
are assumed to hold a fraction $\epsilon_e$ of the internal energy
immediately behind the shock, while the magnetic field is assumed to
hold a fraction $\epsilon_B$ of the internal energy everywhere in the
shocked region.

The spectral emissivity in the comoving frame of the emitting shocked
material is typically approximated as a broken power law (in some
cases the more accurate functional form of the synchrotron emission is
used, e.g., Wijers \& Galama 1999, Granot \& Sari 2002). Most
calculations of the light curve assume emission from an infinitely
thin shell, which represents the shock front (some integrate over the
volume of the shocked fluid taking into account the appropriate radial
profile of the flow, e.g., Granot \& Sari 2002; see
Figure~\ref{fig:spectrum}). One also needs to account for the
different arrival times of photons to the observer from emission at
different lab frame times and locations relative to the line of sight,
as well as the relevant Lorentz transformations of the emission into
the observer frame. SSC is included in some (but not all) works, even
though it can also effect the synchrotron emission through the
enhanced radiative cooling of the electrons.

\section{Afterglow images}
\label{images}

In order to calculate the afterglow image, or how the afterglow would
appear on the plane of the sky if it were angularly resolved, we need
to specify the dynamics, in addition to an emission model, here
assumed to be synchrotron radiation. As discussed in
Sect.~\ref{dynamics}, at early times before the jet break time in the
afterglow light curve the dynamics may be reasonably approximated as
part of a spherical flow that is described by the self-similar
Blandford \& McKee (1976) solution. Here we shall concentrate on this
stage, for which the afterglow image is also self-similar.

\begin{figure*}
%\begin{center}
\epsfxsize=12cm \epsfbox{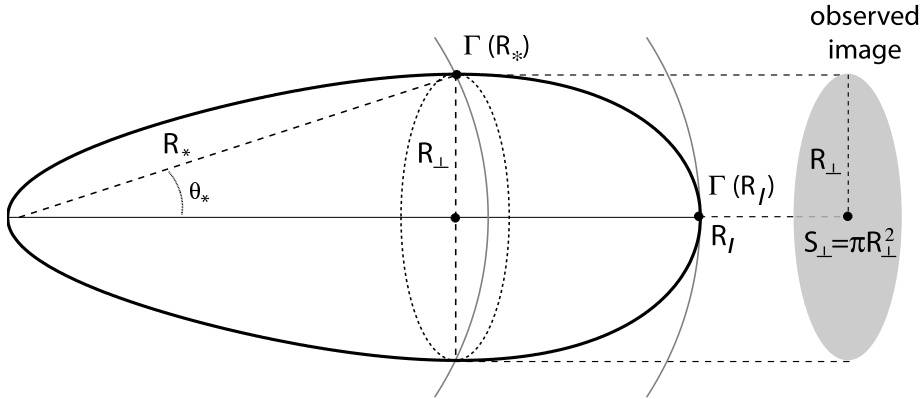}
%\end{center}
\caption{Schematic illustration of the equal arrival time surface
  (thick black line), namely, the surface from where the photons
  emitted at the shock front arrive at the same time at the observer
  (on the far right-hand side). The maximal lateral extent of the
  observed image, $R_\perp$, is located at an angle , where the shock
  radius and Lorentz factor are $R_*$ and $\Gamma_* = \Gamma_{\rm
  sh}(R_*)$, respectively. The area of the image on the plane of the
  sky is $S_\perp = \pi R_\perp^2$. The shock Lorentz factor,
  $\Gamma_{\rm sh}$, varies with radius $R$ and angle $\theta$ from
  the line of sight along the equal arrival time surface. The maximal
  radius $R_l$ on the equal arrival time surface is located along the
  line of sight. If, as expected, $\Gamma_{\rm sh}$ decreases with
  $R$, then $\Gamma_l = \Gamma_{\rm sh}(R_l)$ is the minimal shock
  Lorentz factor on the equal arrival time surface. (from Granot et~al.\
  2005).}
\label{fig:im_diag}
\end{figure*}

During the self-similar spherical evolution stage (before the jet
break time, for a jet), the afterglow image has circular symmetry
around the line of sight (where the surface brightness depends only on
the distance from the center of the image), and is confined within a
circle on the sky with a radius
\begin{equation}
\frac{R_\perp}{10^{16}\,{\rm cm}} = \left\{\matrix{
  3.91\left(\frac{E_{52}}{n_0}\right)^{1/8}\left(\frac{t_{\rm
      days}}{1+z}\right)^{5/8} & \ \ (k=0)\ , \cr & \cr
  2.39\left(\frac{E_{52}}{A_*}\right)^{1/4}\left(\frac{t_{\rm
      days}}{1+z}\right)^{3/4} & \ \ (k=2)\ ,}\right.
\end{equation}
(see Figure~\ref{fig:im_diag}).

This corresponds to an angular radius of
\begin{equation}
\frac{R_\perp}{d_A} = \left\{\matrix{
\frac{1.61\,\mu{\rm
as}}{d_{A,27.7}}\left(\frac{E_{52}}{n_0}\right)^{1/8}
\left(\frac{t_{\rm days}}{1+z}\right)^{5/8} & \ \ (k=0)\ , \cr & \cr
\frac{0.98\,\mu{\rm
as}}{d_{A,27.7}}\left(\frac{E_{52}}{A_*}\right)^{1/4}
\left(\frac{t_{\rm days}}{1+z}\right)^{3/4} & \ \ (k=2)\ ,} \right.
\end{equation}
where $d_A(z) = 10^{27.7}d_{A,27.7}\,{\rm cm}$~\footnote{For a
standard cosmology ($\Omega_M = 0.27$, $\Omega_\Lambda = 0.73$, $h =
0.71$) $d_A(z)$ has a maximum value of $5.45\times 10^{27}\,$cm
($d_{A,27.7} = 1.09$) for $z = 1.64$.}

\begin{figure*}
\begin{center}
\epsfxsize=11cm \epsfbox{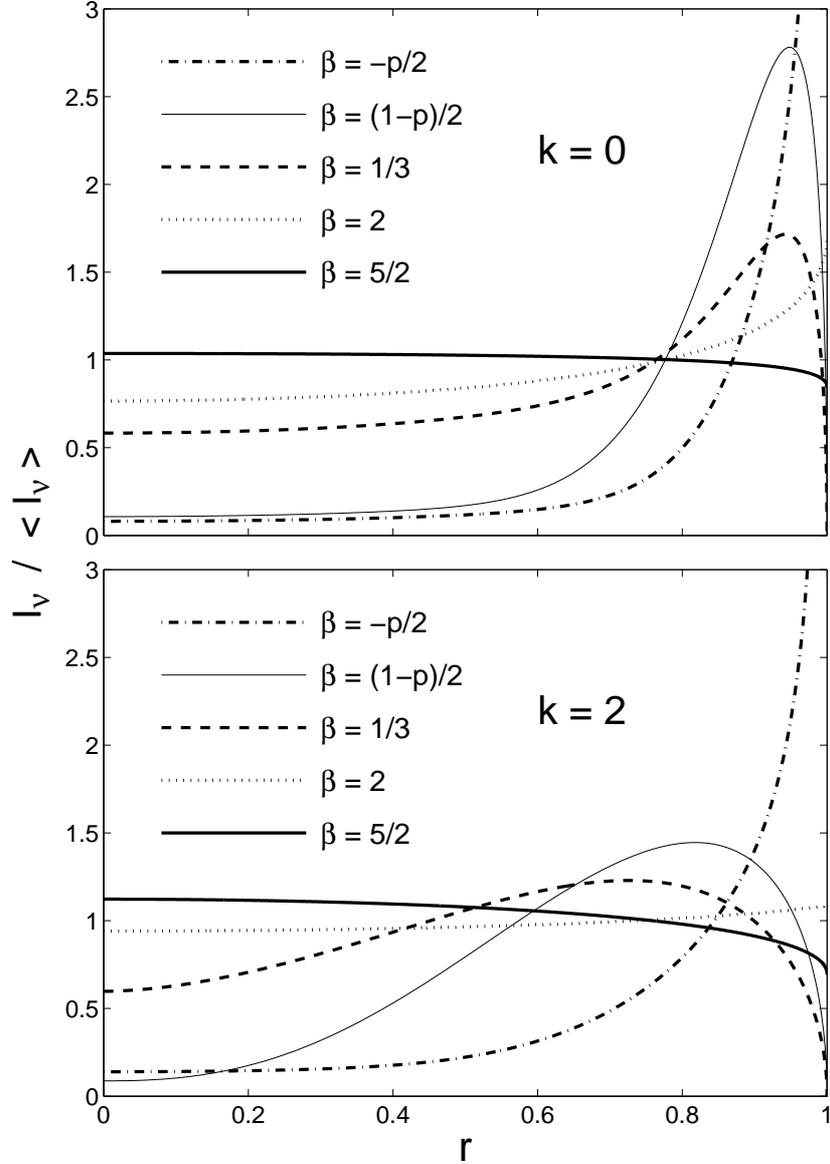}
\end{center}
\caption{The afterglow images for different power-law segments of the
  spectrum, for a uniform ($k = 0$) and wind ($k = 2$) external
  density profile (from Granot \& Loeb 2001), calculated for the
  Blandford \& McKee (1976) spherical self similar solution, using the
  formalism of Granot \& Sari (2002). Shown is the surface brightness,
  normalized by its average value, as a function of the normalized
  distance from the center of the image, $r = R\sin\theta/R_\perp$
  (where $r = 0$ at the center and $r = 1$ at the outer edge). The
  image profile changes considerably between different power-law
  segments of the afterglow spectrum, $F_\nu \propto \nu^\beta$. There
  is also a strong dependence on the density profile of the external
  medium, $\rho_{\rm ext} \propto R^{-k}$.}
\label{fig:afterglow_images}
\end{figure*}

\begin{figure*}
\begin{center}
\epsfxsize=4.1cm \epsfbox{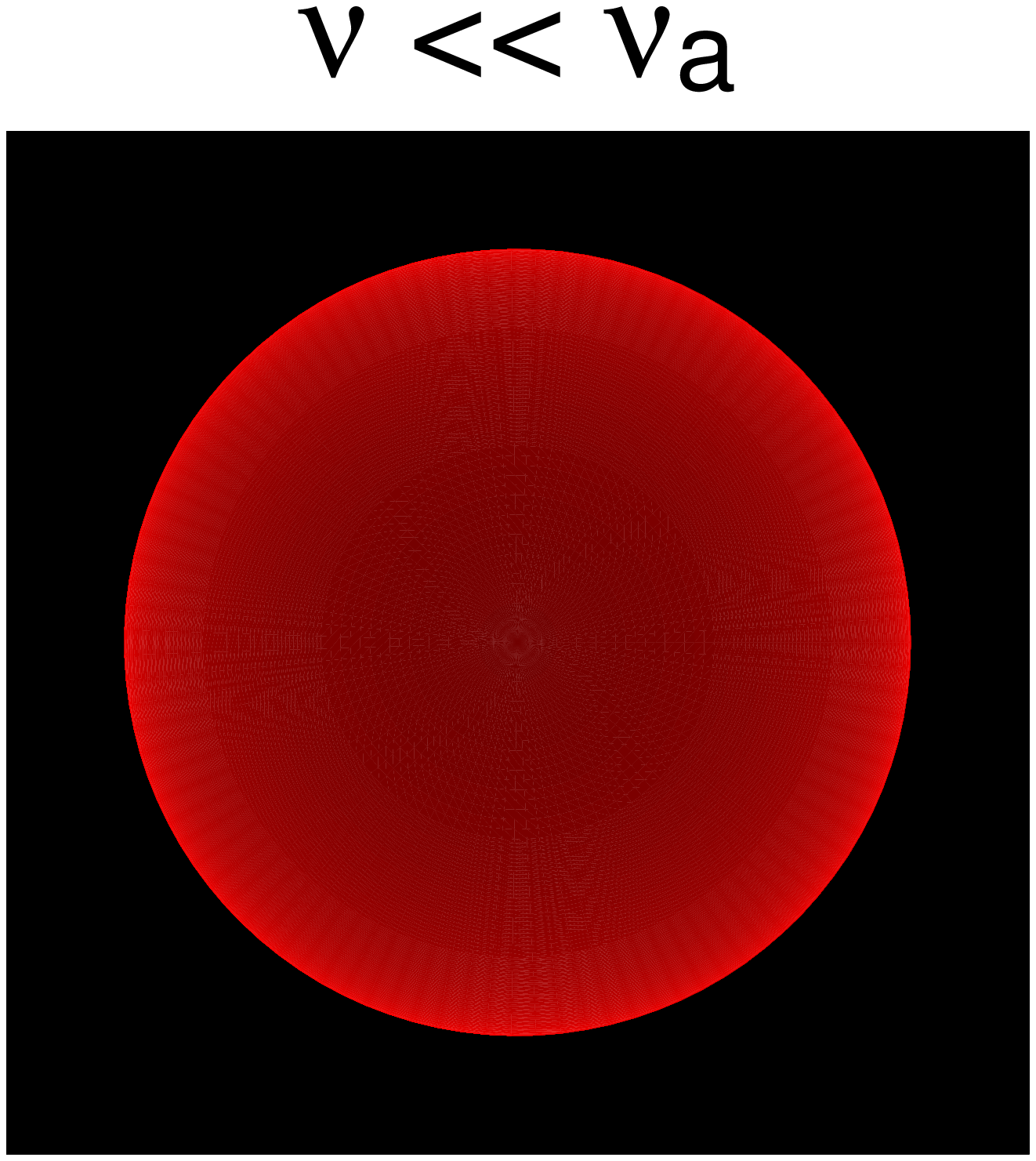} 
\epsfxsize=4.1cm \epsfbox{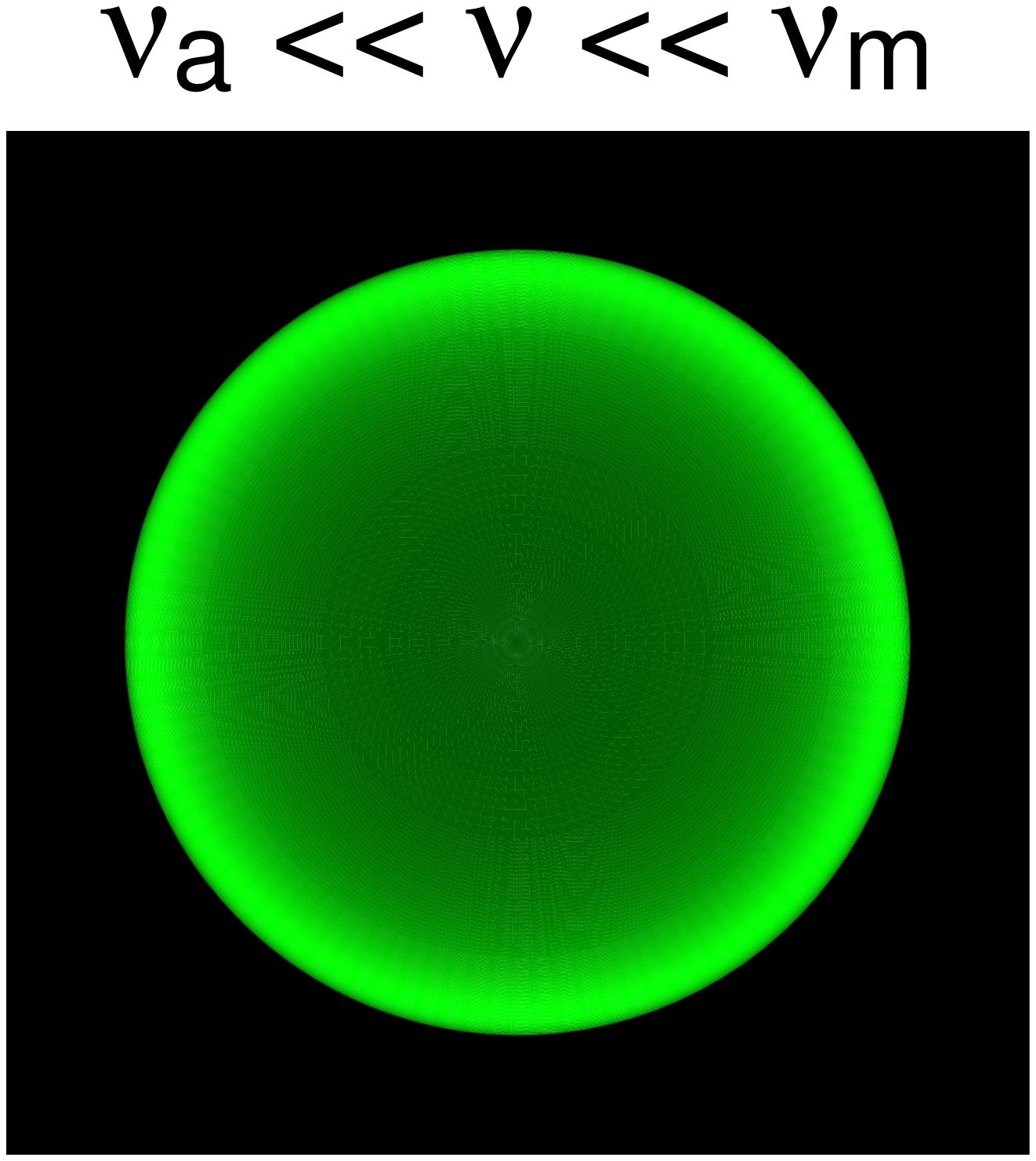} 
\epsfxsize=4.1cm \epsfbox{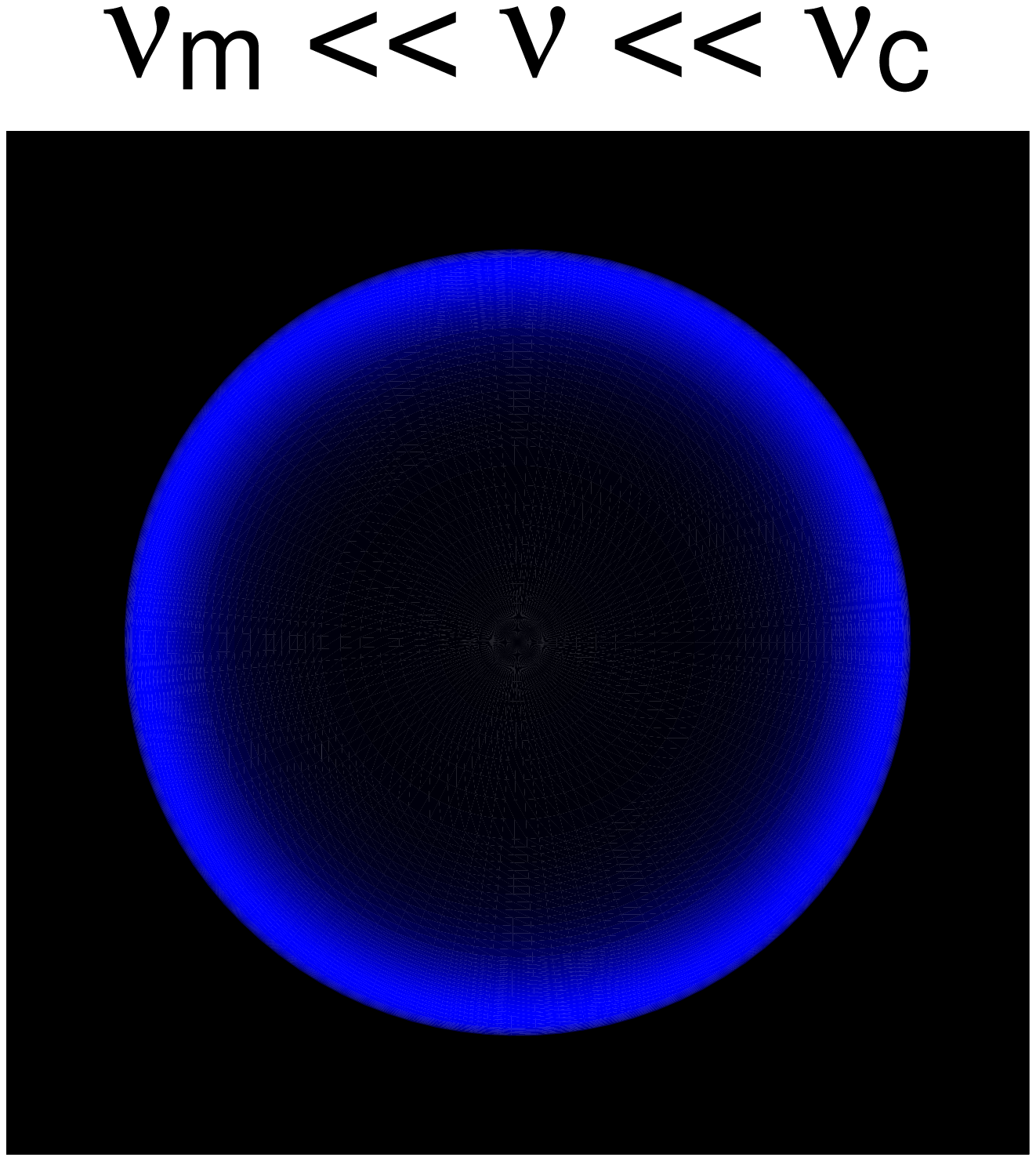}
\end{center}
\caption{An illustration of the expected afterglow image on the plane§
  of the sky, for three different power-law segments of the spectrum
  (from Granot et~al.\ 1999a,b), assuming a uniform external
  density and the Blandford McKee (1976) self-similar solution. The
  image is more limb brightened at power-law segments that correspond
  to higher frequencies.\vspace{-0.14cm}}
\label{fig:afterglow_images_3}
\end{figure*}

More generally, the afterglow image size during the self-similar
spherical stage scales with the observed time as $R_\perp \propto
t^{(5-k)/2(4-k)}$. The image size grows super-luminally with an
apparent expansion velocity of $\Gamma_{\rm sh}(R_*)c$. The expected
afterglow images during this self-similar regime are shown in
Figures~\ref{fig:afterglow_images} and \ref{fig:afterglow_images_3}.
The normalized surface brightness profile within the afterglow image
is independent of time due to the self-similar dynamics, and changes
only between the different power-law segments of the synchrotron
spectrum, and for different external density profiles. The image
becomes increasingly limb-brightened at higher frequencies, and for
smaller values of $k$ (Granot \& Loeb 2001, Granot 2008).

Below the self-absorption frequency, the specific intensity (surface
brightness) represents the Rayleigh-Jeans portion of a black-body
spectrum with the blue-shifted effective temperature of the electrons
at the corresponding radius along the front side of the equal arrival
time surface of photons to the observer ($R_* \leq R \leq R_l$ in
Figure~\ref{fig:im_diag}). Above the cooling break frequency the
emission originates from a very thin layer behind the shock front,
where the electrons whose typical synchrotron frequency is close to
the observed frequency have not yet had enough time to significantly
cool due to radiative losses. This results in a divergence of the
surface brightness at the outer edge of the image (Sari 1998, Granot
\& Loeb 2001).

After the jet break time the afterglow image is no longer symmetric
around the line of sight to the central source for a general viewing
angle (which is not exactly along the jet symmetry axis), and its
details depend on the hydrodynamic evolution of the jet (so that in
principle it could be used to constrain the jet dynamics).
Therefore, a realistic calculation of the afterglow image during the
more complicated post-jet break stage requires the use of hydrodynamic
simulations, and still remains to be done.

The afterglow image may be indirectly resolved via gravitational
lensing by a star in an intervening galaxy (along, or close to, our
line of sight to the source). This is because the angular size of the
Einstein radius (i.e., the region of large magnification around the
lensing star) for a typical star at a cosmological distance is
$\sim\,$1$\,\mu$as (hence the name micro-lensing) -- comparable to the
afterglow image size after a day or so. Since the afterglow image size
grows very rapidly with time, different parts of the image sample the
regions of large magnification (close to the point of infinite
magnification just behind the lensing star) with time, and therefore
the overall magnification of the afterglow flux as a function of time
probes the surface brightness profile of the afterglow image. This
results in a bump in the afterglow light curve, which peaks when the
limb-brightened outer part of the image sweeps past the lensing star;
the peak of the bump is sharper when the
afterglow image is more limb-brightened (Granot \& Loeb 2001). 
It has been suggested that an
achromatic bump in the afterglow light curve of GRB\,000301C after
$\sim 4\,$days might have been due to micro-lensing (Garnavich et~al.\
2000). If this interpretation is true, then the shape of the
bump in the afterglow light curve requires a limb-brightened 
afterglow image, in agreement with theoretical expectations 
(Gaudi et~al.\ 2001).

\section{The cause of the jet break}
\label{jet_break}

The jet break in the afterglow light curve has been argued to be the
combination of (i) the edge of the jet becoming visible, and (ii) fast
lateral spreading. Both effects are expected to take place around the
same time, when the Lorentz factor, $\Gamma$, of the jet drops below
the inverse of its initial half-opening angle, $\theta_0$. This can be
understood as follows.

When $\Gamma$ drops below $\theta_0^{-1}$, the edge of the jet becomes
visible. This is because relativistic beaming limits the region from
which a significant fraction of the emitted radiation reaches the
observer to within an angle of $\sim \Gamma^{-1}$ around the line of
sight ($\theta \lesssim \Gamma^{-1}$). Once the edge of the jet
becomes visible, then, if there is no significant lateral spreading,
only a small fraction, $(\Gamma\theta_{\rm j})^2 < 1$, of the visible region
is occupied by the jet and, as a result, there would be ``missing''
contributions to the observed flux compared to a spherical flow. This
would cause a steepening in the light curve, i.e., a jet break, where
the temporal decay index asymptotically increases by $\Delta\alpha =
(3-k)/(4-k)$ (since the fraction of the visible region occupied by the
jet is $(\Gamma\theta_{\rm j})^2 \sim (t/t_{\rm j})^{-(3-k)/(k-4)}$).

\begin{figure*}
%\vspace{0.1cm}
%\hspace{0.10cm}
\begin{center}
\epsfxsize=0.6\textwidth \epsfbox{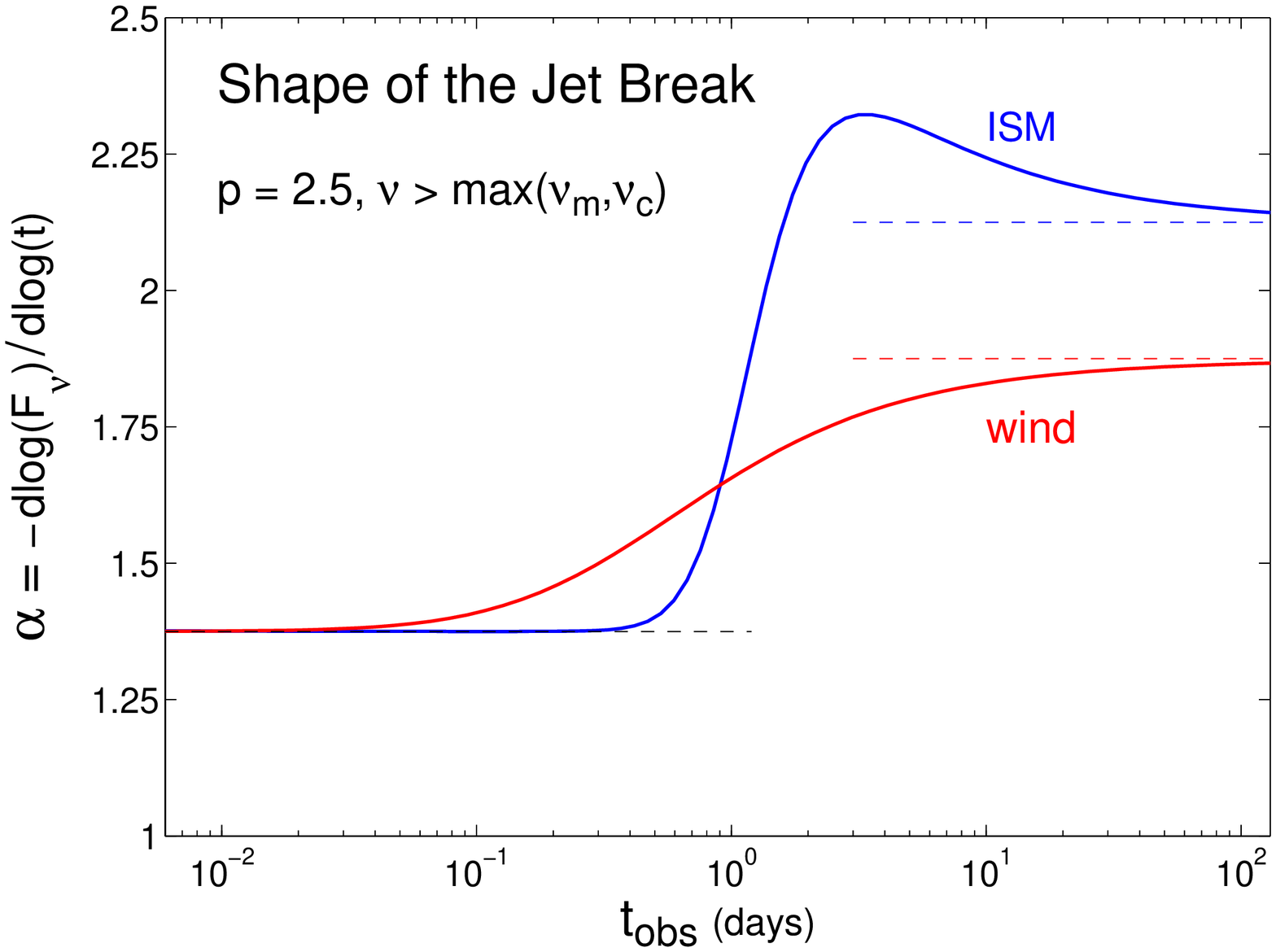}\\
\vspace{0.5cm}
\epsfxsize=0.6\textwidth \epsfbox{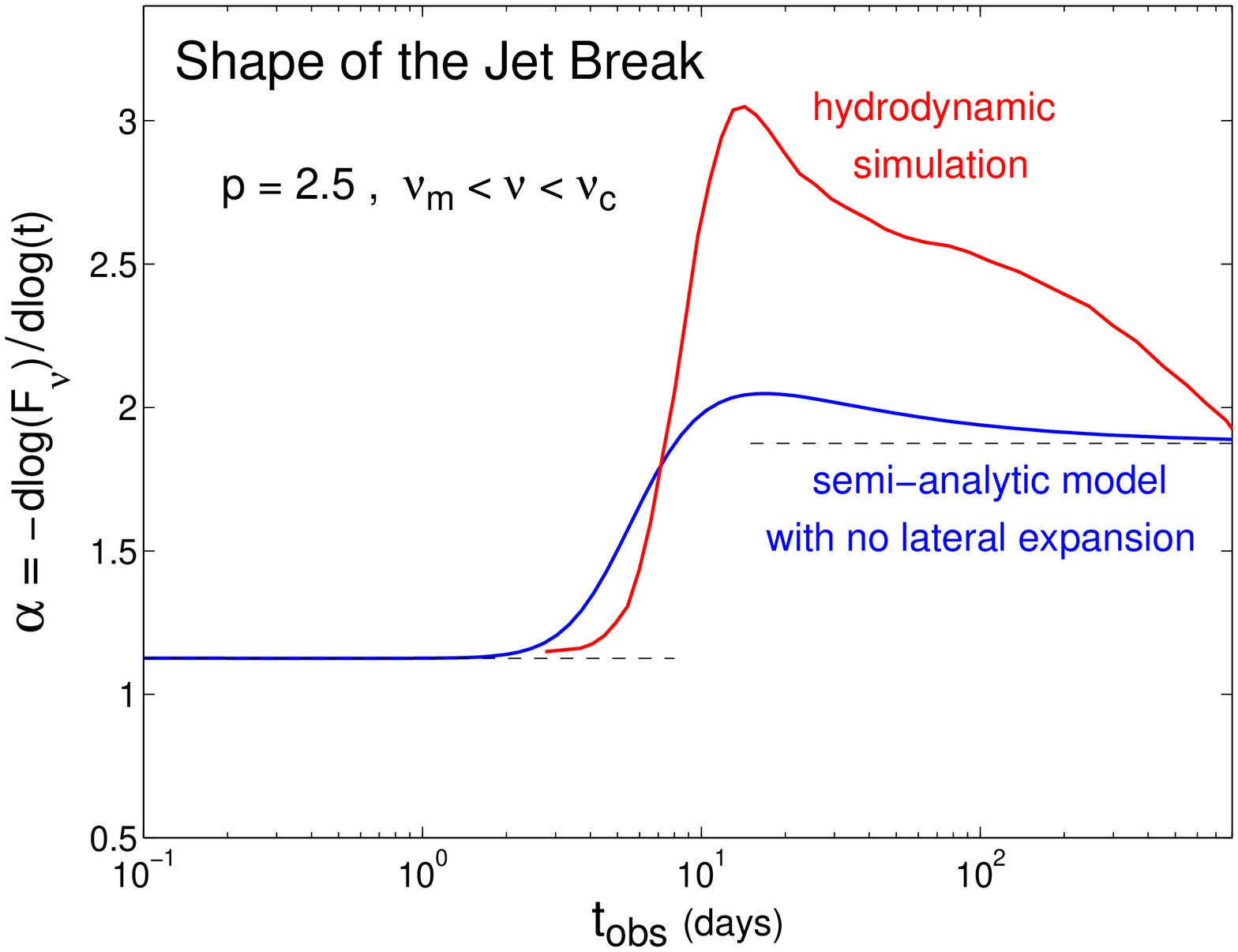} \\
\end{center}
\caption{The temporal decay index $\alpha$ as a function of the
  observed time (in days) across the jet break in the light curve, for
  $p = 2.5$. {\it Top panel}: results in the spectral range $\nu >
  \max(\nu_m,\nu_c)$ using a semi-analytic model with no lateral
  spreading (Granot 2005), for a uniform ($k = 0$, $n_{\rm ext} =
  1\,{\rm cm^{-3}}$) and wind ($k = 2$, $A_* = 1$) external density
  profile, with $\theta_0 = 0.1$ and $E_{\rm k,iso} = 2\times
  10^{53}\,$erg.  {\it Bottom panel}: results for the spectral range
  $\nu_m < \nu < \nu_c$, for $\theta_0 = 0.2$ and a uniform density
  ($k = 0$, $n_{\rm ext} = 1\,{\rm cm^{-3}}$, $E_{\rm k,iso} =
  10^{52}\,$erg); the figure
  compares the result of a semi-analytic model (Granot
  2005) to those of a hydrodynamic simulation (Granot et~al.\ 2001). In
  both panels the dashed lines show the asymptotic values of $\alpha$
  before and after the jet break, for a uniform jet with no lateral
  spreading, for which $\Delta\alpha = (3-k)/(4-k)$.}
\label{fig:jet_break}
\end{figure*}

When $\Gamma$ drops below $\theta_0^{-1}$, the center of the jet comes
into causal contact with its edge, and the jet can in principle start
to expand sideways significantly. It has been argued that at this
stage it would indeed start to expand sideways rapidly, at close to
the speed of light in its own rest frame. In this case, during the
rapid lateral expansion phase the jet opening angle grows as $\theta_{\rm j}
\sim \Gamma^{-1}$ and exponentially with radius (see Sect.~\ref{dynamics}). 
This causes the energy per unit solid angle, $\mathcal{E}$, in the jet to
drop with observed time, and the Lorentz factor to decrease faster as
a function of observed time; this gives rise to a steepening in
the afterglow light curve compared to a spherical flow (for
which $\mathcal{E}$ remains constant and $\Gamma$ decreases more
slowly with the observed time). However, in this case a large fraction of
the visible region remains occupied by the jet (since $\Gamma\theta_{\rm j}$
remains $\sim 1$), so that the first cause for the jet break (the edge
of the jet becoming visible, and the ``missing'' contributions from
outside the edge of the jet) is no longer important. Therefore,
for fast lateral spreading, the jet break occurs predominantly both as
a result of the energy per solid angle $\mathcal{E}$ decreasing with
time and the Lorentz factor decreasing with observed time faster than
for a spherical flow.

It is important to keep in mind, however, that numerical studies show
that the lateral spreading of the jet is very modest as long as it is
relativistic (see Sect.~\ref{dynamics}). This implies that lateral
spreading cannot play an important role in the jet break, and the
predominant cause of the jet break is the ``missing'' contribution
from outside of the jet, once its edge becomes visible.

A potential problem with this description is that if the jet half-opening
angle remains roughly constant, $\theta_{\rm j} \approx \theta_0$, the
asymptotic change in the temporal decay index is only $\Delta \alpha =
3/4$ for a uniform external medium ($k = 0$) or even smaller for a
wind ($\Delta\alpha = 1/2$ for $k = 2$), while the values inferred
from observations are in most cases larger (see Figure 3 of Zeh et~al.\
2006). This apparent discrepancy may be reconciled as follows.
While the {\it asymptotic} steepening is indeed $\Delta \alpha =
(3-k)/(4-k)$ when lateral expansion is negligible, the value of the
temporal decay index $\alpha$ (where $F_\nu \propto t^{-\alpha}$)
initially overshoots its asymptotic value. Since the temporal baseline
that is used to estimate the post-jet break temporal decay
index $\alpha_2$ is typically no more than a factor of several in time
after the jet break time\footnote{This is usually because the flux
becomes too dim to detect above the host galaxy, or since a supernova
component becomes dominant in the optical.}, $t_{\rm j}$, the value of
$\alpha$ during this time is {\it larger} than its asymptotic value,
$\alpha_2$.  This causes the value of $\Delta\alpha$ that is inferred
from observations to be larger than its true asymptotic value.

The overshoot in the value of $\alpha$ just after the jet break time
can be seen in Figure~\ref{fig:jet_break}, and is much more pronounced
in the light curves calculated using the jet dynamics from a
hydrodynamic simulation, compared to the results of a simple
semi-analytic model. This overshoot occurs because the afterglow image
is limb-brightened (see Figure~\ref{fig:afterglow_images}) and
therefore its brightest part -- its outer edges -- is the first region
whose contribution to the observed flux is ``missed'' as the edge of
the jet becomes visible. The overshoot increases for increasingly
limb-brightened afterglow images (e.g.,, for $\nu > \max[\nu_m,\nu_c]$
in the upper panel of Figure~\ref{fig:jet_break} compared to $\nu_m <
\nu < \nu_c$ in the lower panel of Figure~\ref{fig:jet_break}). For a
wind density ($k = 2$) the limb-brightening is smaller compared to a
uniform density ($k = 0$), at the same power-law segment of the
spectrum (see Figure~\ref{fig:afterglow_images}), and the Lorentz
factor $\Gamma$ decreases more slowly with the observed time. Because
of this no overshoot is seen in the semi-analytic model shown in the
upper panel of Figure~\ref{fig:jet_break} for a wind density profile
($k = 2$), and the jet break is smoother and extends over a larger
factor in time. The asymptotic post-jet break value of the temporal
decay index ($\alpha_2$) is approached only when the part of the image
of the corresponding spherical flow that is occupied by the jet covers
only the relatively uniform central part of the image, and not its
brighter outer edge.

The jet break in light curves calculated from hydrodynamic simulations
is sharper than in semi-analytic models (where the emission is taken
to be from a 2D surface -- usually a section of a sphere within a
cone). In semi-analytic models the jet break is sharpest with no
lateral expansion, and becomes more gradual with faster assumed lateral
expansion. For example, in the lower
panel of Figure~\ref{fig:jet_break}, where the viewing angle is along
the jet axis and the external density is uniform, most of the change
in the temporal decay index $\alpha$ occurs over a factor of $\sim 2$
in time for the numerical simulation, and over a factor of $\sim 3$ in
time for the semi-analytic model (which assumes no lateral expansion;
the jet break would be more gradual with lateral expansion). For both
types of models, the jet break is more gradual and occurs at a
somewhat later time for viewing angles further away from the jet
symmetry axis but still within its initial opening angle, although
this effect is somewhat more pronounced in semi-analytic models
(Granot et~al.\ 2001, Rossi et~al.\ 2004).

\section{Polarization: afterglow and prompt}
\label{pol}

\begin{figure*}
%\vspace{0.1cm}
%\hspace{0.10cm}
\epsfxsize=5.9cm \epsfbox{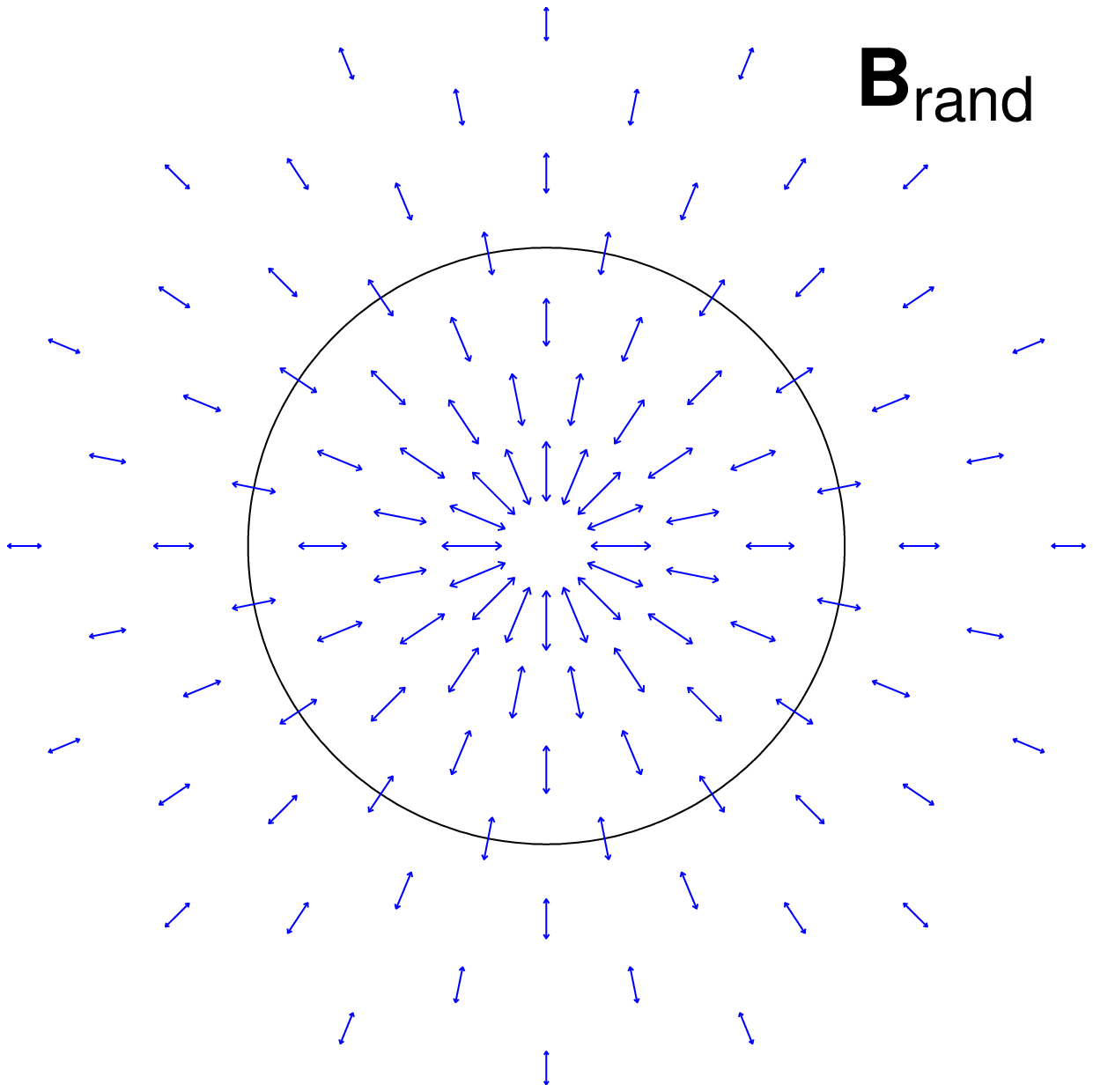}
\hspace{0.53cm}
\epsfxsize=5.9cm \epsfbox{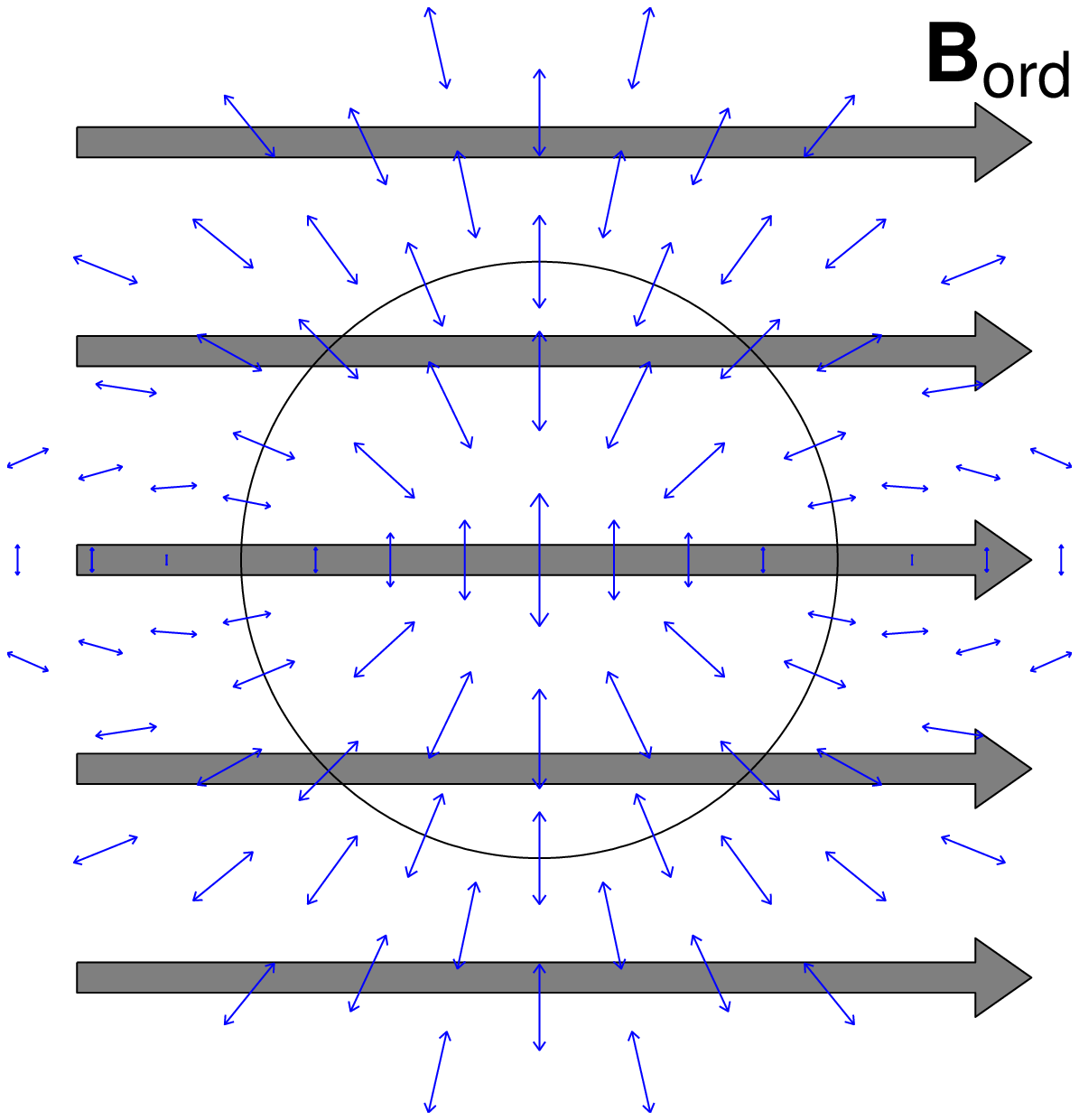} 
\\
\caption{The predicted polarization map for synchrotron emission from 
a thin spherical ultra-relativistic shell expanding with a Lorentz
factor $\Gamma \gg 1$. The double-sided arrows show the direction of
the linear polarization (the wave electric vector), while their length
depends monotonically on the polarized intensity (in a non-trivial
way, for display purposes). The circle indicates an angle of
$1/\Gamma$ around the line of sight, which is the region responsible
for most of the observed flux. The {\it left panel} is for a magnetic
field that is random within the plane of the shell (normal to the
radial direction), for which the polarization direction always points
at the center of the image (corresponding to the line of sight to the
center of the spherical shell), where the polarization vanishes (due
to symmetry consideration). The {\it right panel} is for an ordered
magnetic field within the plane of the shell that is coherent over
angular scales $\gg 1/\Gamma$ (Granot \& K\"onigl 2003). In this case
the direction of the ordered magnetic field clearly breaks the
symmetry around the center of the image, resulting in a large net
polarization. For simplicity, the map is for a constant emission
radius, rather than for a constant photon arrival time.}
\label{fig:plo_ord-rand}
\end{figure*}

{\bf Afterglow:} Linear polarization at the level of a few percent has
been detected in the optical and NIR afterglow of several GRBs (see
Covino et~al.\ 2004 and references therein). This was considered to be a
confirmation that synchrotron radiation is the dominant emission
mechanism in the afterglow. The synchrotron emission from a fluid
element with a locally uniform magnetic field is linearly polarized in
the direction perpendicular to the projection of the magnetic field
onto the plane normal to the wave vector. Since the source moves
relativistically, aberration of light (see Sect.~\ref{beaming}) must be
accounted for when calculating the observed local direction of
polarization. Figure~\ref{fig:plo_ord-rand} shows the predicted local
polarization map from emission by an ultra-relativistic expanding
shell, for two different magnetic field structures: a magnetic field
that is random within the plane normal to the radial direction ({\it
left panel}) as could be expected from a shock produced magnetic field
(e.g., Medvedev \& Loeb 1999), and an ordered magnetic field normal to
the radial direction (as could be expected in the prompt emission for
a magnetic field coherent on angular scales $\gg 1/\Gamma_0$ that is
advected from the central source).

Since the afterglow image is almost always never resolved, we can only
measure the average polarization over the whole image. For this
reason, a shock produced magnetic field that is symmetric about the
shock normal will procure no net polarization for a spherical flow
(since the polarization pattern across the image is in this case
symmetric around the center of the image, and the polarization
averages out to zero when summing over the contributions from the
whole image).  For a shock-produced magnetic field one thus needs to
break the spherical symmetry of the flow to produce net
polarization. The simplest and most natural way of doing this is
considering a jet, or narrowly collimated outflow (e.g., Sari 1999,
Ghisellini \& Lazzati 1999). In this picture a jet geometry together
with a line of sight that is not along the jet symmetry axis (but
still within the jet aperture, in order to see the prompt GRB) is
needed to break the symmetry of the afterglow image around
our line of sight.

For a uniform jet (the UJ model) this predicts two peaks in the
polarization light curve around the jet break time $t_{\rm j}$, if
$\Gamma\theta_{\rm j} < 1$ decreases with time at $t > t_{\rm j}$ (Ghisellini \&
Lazzati 1999, Rossi et~al.\ 2004), or even three peaks if
$\Gamma\theta_{\rm j} \approx 1$ at $t> t_{\rm j}$ (Sari 1999), where in both
cases the polarization vanishes and reappears rotated by $90^\circ$
between adjacent peaks.  The latter is a distinct signature of this
model. For a structured jet (the USJ model), the polarization position
angle is expected to remain constant in time, while the degree of
polarization peaks near the jet break time $t_{\rm j}$ (Rossi et
al. 2004). A similar qualitative behavior is also expected for a
Gaussian jet, or other jet structures with a bright core and dimmer
wings (although there are obviously some quantitative differences).

The different predictions for the afterglow polarization light curves
for different jet structures imply that afterglow polarization
observations could in principle help constrain the jet structure. In
practice, however, the situation is much more complicated, mainly
since the observed polarization depends not only on the jet geometry,
but also on the magnetic field configuration within the emitting
region, which is not well known. For example, an ordered magnetic
field component in the emitting region (e.g., due to a small ordered
magnetic field in the external medium) may dominate the polarized
flux, and therefore the polarization light curves, even if it is
sub-dominant in the emitting region compared to a random (shock
generated) magnetic field component in terms of the total energy in
the magnetic field (Granot \& K\"onigl 2003).  Other models for
afterglow polarization include a magnetic field that is coherent over
patches of a size comparable to that of causally connected regions
(Gruzinov \& Waxman 1999), and polarization that is induced by
microlensing (Loeb \& Perna 1998) or by scintillations in the radio
(Medvedev \& Loeb 1999). 

The last two models involve mechanisms that give a different weight to
emission from different parts of the afterglow image, and thus break
its symmetry around the line of sight for a shock produced magnetic
field (so that the polarization from the different parts of the image
no longer fully cancels out). While microlensing or radio
scintillations are external to the source, there are also mechanisms
intrinsic to the source that can produce a similar effect on the
polarization. An example is a ``patchy shell'', where the energy per
solid angle and the corresponding afterglow brightness vary randomly
with the lateral location within the flow (see
Sect.~\ref{jet_structure}).  As the afterglow shock decelerates and the
angular size of the visible region (of $\lesssim\Gamma^{-1}$ around
the line of sight) increases, new bright and dim regions become
visible and cause correlated variability in the total afterglow flux
(bumps or wiggles in the afterglow light curve) and in its linear
polarization -- both in the degree of polarization and its position
angle (Nakar \& Oren 2004). Soon after this was predicted (Granot \&
K\"onigl 2003),\footnote{A correlated variability in the afterglow
light curve and its polarization is also expected in other scenarios,
such as microlensing, radio scintillation and non-axisymmetric
``refreshed shocks'' or variations in the external density.} such a
correlated variability in the afterglow light curve and its
polarization was reported for two particularly variable optical
afterglows, of GRB 021004 (Rol et~al.\ 2003) and GRB 030329 (Greiner et
al. 2003).

{\bf Prompt Emission:} in the prompt soft gamma-ray emission the
nature of the dominant emission mechanism remains uncertain. Moreover,
it is very difficult to measure the polarization at such photon
energies (hard X-rays or soft gamma-rays), compared to the optical,
NIR, or radio. There have been some claims (Coburn \& Boggs 2003,
Willis et~al.\ 2005, McGlynn et~al.\ 2007, G\"otz et~al.\ 2009) of
detection of a high degree of linear polarization in the prompt
gamma-ray emission of some GRBs (with a particularly high fluence, as
good photon statistics are crucial for such measurements). However,
these claims remain rather controversial (Rutledge \& Fox 2004, Wigger
et~al.\ 2004). A detection of polarization during the prompt emission
phase could nonetheless help constrain the dominant emission
mechanism.

An ordered magnetic field within the outflow on angular scales
$\gtrsim 1/\Gamma$ can produce a high net polarization, of tens of
percent (Granot \& K\"onigl 2003, Granot 2003, Lyutikov et~al.\ 2003),
which is only slightly smaller than the maximal polarization of the
local synchrotron emission, as the polarization vector is more or less
aligned in the regions that contribute the most to the observed flux
(see right panel of Figure~\ref{fig:plo_ord-rand}).  Such a high
polarization can also be produced by synchrotron emission from a shock
generated magnetic field (Waxman 2003, Granot 2003, Nakar et~al.\  
2003) or bulk inverse-Compton scattering of an external photon
field (Shaviv \& Dar 1995, Lazzati et~al.\ 2004). However, these
mechanisms require the line of sight to be near the edge of the jet
(to within an angle of $\lesssim 1/\Gamma$).~\footnote{A significant
change in the brightness of the jet on an angular scale of $\sim
1/\Gamma$ around the line of sight can mimic a line of sight near the
edge of a uniform jet, and produce a similarly high degree of
polarization.}  In order for such a line of sight not to be too rare,
a narrow jet is required: $\theta_0 \lesssim {\rm a\ few\ }
\Gamma^{-1}$. This requirement may be viable for the brightest GRBs, for
which the prompt polarization can be measured, as such bright events
usually correspond to very narrow jets. Nevertheless, statistical
studies over a sample of GRBs, or time resolved polarimetry of
different emission episodes within a single very bright GRB, may help
distinguish between the different possible causes for polarization and
teach us about the dominant emission mechanism, the jet structure, or
the magnetic field configuration within the GRB outflow.

\section{Light curves for off-axis viewing angles}
\label{off-axis_LC}

The observed flux density $F_\nu = dE/dAd\nu dt$ is the energy per
unit area, frequency and time in the direction $\hat{n}_d$ normal to a
surface area $dA$ (at the detector). The differential contribution to
it is $dF_\nu(\hat{n}_d) =
I_\nu(\hat{n})\cos\theta_{sd}\,d\Omega_{sd}$ where $\cos\theta_{sd} =
\hat{n} \cdot \hat{n}_d$, $\hat{n}$ being the local direction from the
relevant part of the source to the observer (or detector),
$d\Omega_{sd} = d\phi_{sd}d\cos\theta_{sd}$ is the differential solid
angle sustained by the contributing portion of the source as viewed
from the observer, and $I_\nu(\hat{n}) = dE/dAd\Omega d\nu dt$ is the
specific intensity (the energy per unit area, time and frequency of
radiation directed within a small solid angle $d\Omega$ centered on
the direction $\hat{n}$). Since in practice almost always $\theta_{sd}
\ll 1$, one can simplify the expression for $F_\nu$ by approximating
$\cos\theta_{sd} \approx 1$. Furthermore, we have $d\Omega_{sd} =
dS_\perp/d_A^2$ where $d_A(z)$ is the angular distance to the source,
and $dS_\perp$ is the differential area in the plane of the sky
(normal to $\hat{n}$) sustained by the source, so that $dF_\nu = I_\nu
dS_\perp/d_A^2$. Let us denote with a subscript $z$ quantities
measured at the cosmological frame of the source.  For an optically
thin source $I_{\nu_z} = \int j_{\nu_z} ds_z$, where $j_{\nu_z} =
dE_z/dV_zd\Omega_z d\nu_z dt_z$ is the emitted energy per unit volume,
solid angle, frequency and time, while $ds_z$ is the differential path
length along the trajectory of a photon that reaches the observer at
the time $t_{\rm obs}$ when $F_\nu$ is measured. Since $I_\nu/\nu^3$,
$j_\nu/\nu^2$ and $ds/\nu$ are Lorentz invariant (Rybicki \& Lightman
1979), we have $I_\nu = (\nu/\nu_z)^3I_{\nu_z} =
(1+z)^{-3}\int j_{\nu_z}ds_z$. In addition, since $d_A = d_L/(1+z)^2$,
where $d_L(z)$ is the luminosity distance, then $dF_\nu =
j_{\nu_z}dV_z\,(1+z)/d_L^2$, where $dV_z = dS_\perp ds_z$ is the
volume element in the source cosmological frame. Here $j_{\nu_z} =
[\Gamma(1-\hat{n}\cdot\vec{\beta})]^{-2}\,j'_{\nu'}$ is measured in
the source frame, while $j'_{\nu'}$ is measured in the (comoving) rest
frame of the emitting material, which moves at a velocity
$\vec{\beta}c$ in the source frame. Altogether, this gives
\begin{equation}\label{F_nu1}
F_\nu(t_{\rm obs},\hat{n}) = \frac{(1+z)}{d_L^2(z)}\int d^4x\,\delta
\left(t_z-\frac{t_{\rm obs}}{1+z}-\frac{\hat{n}\cdot\vec{r}}{c}\right)
\frac{j'_{\nu'}}{\Gamma^2(1-\hat{n}\cdot\vec{\beta})^2}\ ,
\end{equation}
where $\nu' = (1+z)\Gamma(1-\hat{n}\cdot\vec{\beta})\nu$, and $t_z$ is
the coordinate time at the source's cosmological frame. Since
$d^4x = dt_zdV_z = dt_zdS_\perp ds_z = dt_zdS_\perp ds'(\nu_z/\nu') =
dt_zdV'/\Gamma(1-\hat{n}\cdot\vec{\beta})$ and $4\pi j'_{\nu'}dV' =
dL'_{\nu'} = 4\pi(dE'/d\Omega'd\nu'dt')$ is the differential of the
isotropic equivalent spectral luminosity in the comoving frame,
eq.~\ref{F_nu1} can be rewritten as
\begin{equation}\label{F_nu2}
F_\nu(t_{\rm obs},\hat{n}) = \frac{(1+z)}{4\pi d_L^2(z)}\int
dt_z\,\delta \left(t_z-\frac{t_{\rm obs}}{1+z}-\frac{\hat{n}\cdot\vec{r}}{c}\right)
\int \frac{dL'_{\nu'}}{\Gamma^3(1-\hat{n}\cdot\vec{\beta})^3}\ .
\end{equation}

This result can be intuitively understood as follows. For a point
source of luminosity $L$ we have $F \equiv L/4\pi d_L^2$ by the
definition of the luminosity distance, and thus $F_\nu = dF/d\nu =
(dL/d\nu_z)(d\nu_z/d\nu)/4\pi d_L^2 = L_{\nu_z}(1+z)/4\pi d_L^2$,
where $L_{\nu_z} = 4\pi(dE_z/d\Omega_z d\nu_z dt_z) =
(dE_z/dE')(d\Omega'/d\Omega_z)(d\nu'dt'/d\nu_zdt_z)L'_{\nu'}$
$=(\nu_z/\nu')^3L'_{\nu'} =
L'_{\nu'}/\Gamma^3(1-\hat{n}\cdot\vec{\beta})^3$ and we have used the
relations $d\Omega'/d\Omega_z = \mathcal{D}^2 = (\nu_z/\nu')^2$ and
$d\nu' dt' = d\nu_z dt_z$ derived in Sect.~\ref{beaming}. This is the
basic result for a point source, and the flux from a source of finite
size can be obtained by dividing it into a large number of small
regions that may be treated as individual point sources (in the sense
that $j_{\nu_z} = [\Gamma(1-\hat{n}\cdot\vec{\beta})]^{-2}\,j'_{\nu'}$
does not vary a lot within such a region) and summing over their
contributions, as manifested in eq.~\ref{F_nu2}.

In this section we are particularly interested in the dependence of
the observed light curve, $F_\nu(t_{\rm obs},\hat{n})$, with viewing
angle $\theta_{\rm obs}$ relative to the symmetry axis of the jet,
that points at some direction $\hat{n}_{\rm j}$, so that $\cos\theta_{\rm
obs} = \hat{n} \cdot \hat{n}_{\rm j}$. For a jet that possesses such axial
symmetry, the observed flux depends only on $\theta_{\rm obs}$ and not
on the azimuthal viewing angle $\phi_{\rm obs}$, so that $F_\nu =
F_\nu(t_{\rm obs},\theta_{\rm obs})$. In order to gain some intuition
for how the afterglow light curve varies with viewing angle it is
useful to first consider the simplest case of a point source moving at
an angle of $\theta_{\rm obs} \ll 1$ from the line of sight with
$\Gamma \gg 1$ where $\Gamma \propto R^{-m/2}$. In this case we can
write
\begin{eqnarray}
F_\nu(t_{\rm obs},\theta_{\rm obs}) \approx \frac{1+z}{4\pi
d_L^2}\mathcal{D}^3L'_{\nu'}(R)\ ,\quad
\mathcal{D}(R,\theta_{\rm obs})\equiv\frac{\nu_z}{\nu'} \approx 
\frac{2\Gamma(R)}{1+[\Gamma(R)\theta_{\rm obs}]^2}\ ,\quad
\\ \label{t_obs_R_theta_obs}
t_{\rm obs}(R,\theta_{\rm obs}) \approx \frac{(1+z)R}{2c\Gamma^2(R)}
\left(\frac{1}{1+m}+[\Gamma(R)\theta_{\rm obs}]^2\right)\ ,
\quad\quad\quad\quad\quad\quad
\end{eqnarray}
where one uses the appropriate value of $R(t_{\rm obs},\theta_{\rm
obs})$ according to eq.~\ref{t_obs_R_theta_obs}.  For a given value of
$\Gamma$, the Doppler factor $\mathcal{D}$ is roughly constant ($\sim
\Gamma$) for viewing angles within the ``beaming cone'' of
half-opening angle $1/\Gamma$ around the direction of motion (i.e., for
$\theta_{\rm obs} \lesssim 1/\Gamma$) and rapidly decreases as
$\mathcal{D} \propto \theta_{\rm obs}^{-2}$ for viewing angles outside
the beaming cone ($\theta_{\rm obs} > 1/\Gamma$). It is convenient to
compare the observed flux density $F_\nu$ at different viewing angles
$\theta_{\rm obs}$ and observed times $t_{\rm obs}$, that originate
from the same emission radius $R$ (Granot et~al.\ 2002). In particular,
$F_\nu (\theta_{\rm obs} > 0)$ with $F_\nu (\theta_{\rm obs} = 0)$,
\begin{eqnarray}\nonumber
F_\nu(t_{\rm obs}, \theta_{\rm obs}) \approx a^3 F_{\nu/a}(bt_{\rm
obs},0)\ ,\quad a \equiv 
\frac{\mathcal{D}(R,\theta_{\rm obs})}{\mathcal{D}(R,0)}
\approx \frac{1}{1+[\Gamma(R)\theta_{\rm obs}]^2}\ ,
\\
b \equiv \frac{t_{\rm obs}(R,0)}{t_{\rm obs}(R,\theta_{\rm obs})} 
\approx \frac{1}{1+(1+m)[\Gamma(R)\theta_{\rm obs}]^2} \sim a\ .
\quad\quad
\end{eqnarray}
If the source decelerates, as expected during the afterglow ($m >
0$), then at early times $\Gamma\theta_{\rm obs} \gg 1$ and we have
$t_{\rm obs} \approx (1+z)\theta_{\rm obs}^2R/2c \propto R$ and $a
\approx (1+m)b \approx [\Gamma(R)\theta_{\rm obs}]^{-2} \propto R^m
\propto t_{\rm obs}^m$. For an on-axis flux density of 
$F_\nu(0,t_{\rm obs}) \propto \nu^{-\beta}t_{\rm obs}^{-\alpha}$, this
implies that $F_\nu(\theta_{\rm obs},t_{\rm obs}) \propto
\nu^{-\beta}t_{\rm obs}^{-\alpha+m(3+\beta-\alpha)}$. If the  
jet is assumed to be non-expanding one can thus approximate
$\Gamma(R)$ at the center of the jet by the expression for a spherical
flow, for which $m = 3-k$ (see Sect.~\ref{dynamics}), so that for a
stellar wind environment ($k = 2$), the flux rises as $t_{\rm
obs}^{3+\beta-2\alpha} \sim t_{\rm obs}^1 - t_{\rm obs}^2$ (for $\beta
\sim 1$ and $\alpha \sim 1 - 1.5$), while for a uniform external
medium ($k = 0$), the flux rises much more steeply as $t_{\rm
obs}^{9+3\beta-4\alpha} \sim t_{\rm obs}^6 - t_{\rm obs}^8$. At late
times when the beaming cone widens enough and engulfs the line of
sight ($\Gamma\theta_{\rm obs} < 1$), the off-axis light curve
approaches the on-axis one, since for $\Gamma\theta_{\rm obs} \ll 1$,
$a \approx b \approx 1$ and, as a result, 
$F_\nu(\theta_{\rm obs},t_{\rm obs}) \approx F_\nu(0,t_{\rm obs})$.

\begin{figure*}
%\vspace{0.1cm}
%\hspace{0.10cm}
\epsfxsize=6.1cm \epsfbox{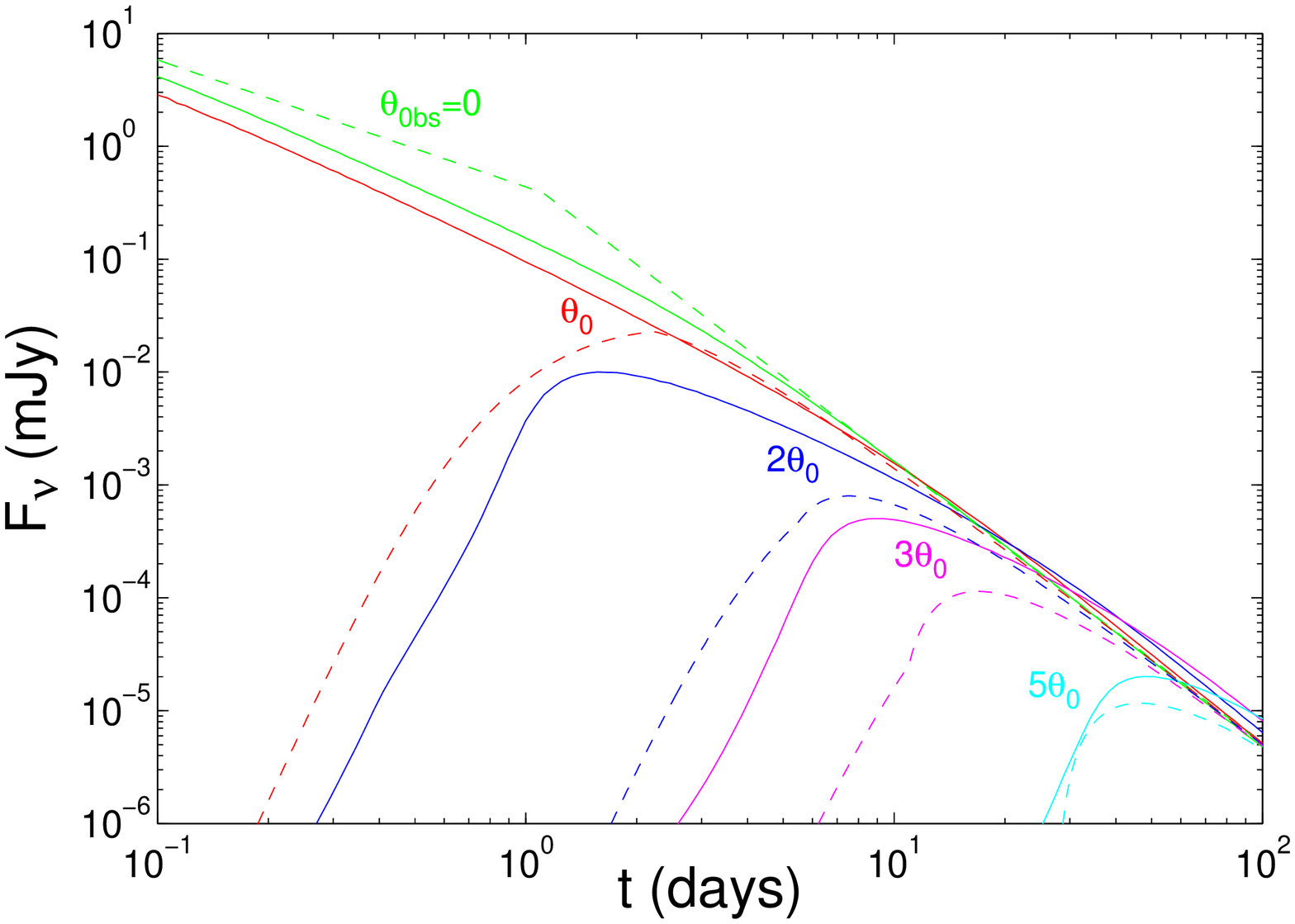}
\hspace{0.5cm}
\epsfxsize=5.8cm \epsfbox{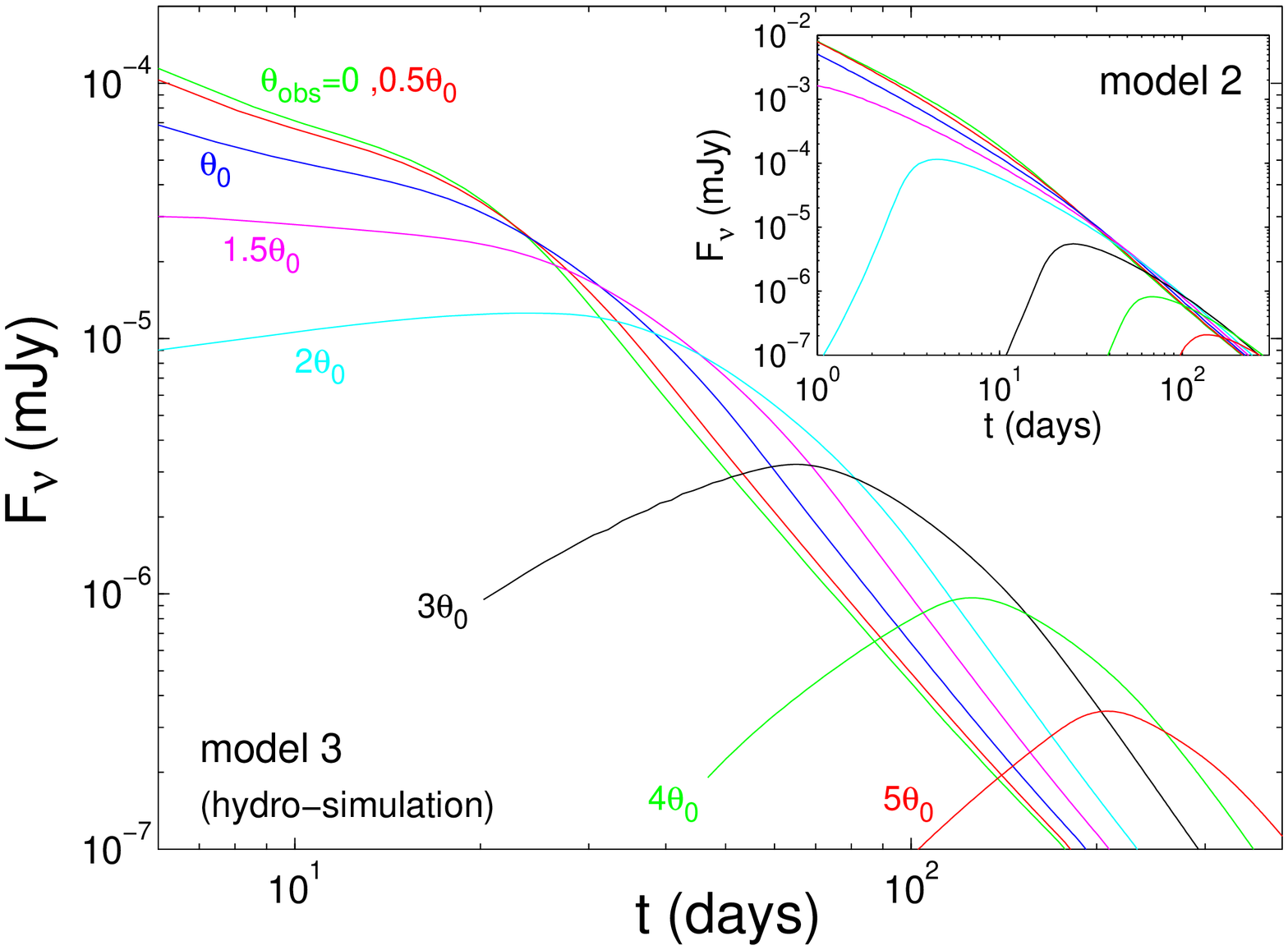} 
\\
\caption{Afterglow light curves for different viewing angles 
(from Granot et~al.\ 2002): {\it Left panel}: comparing a simple
analytic model featuring a point source along the jet axis (model 1;
dashed lines) and a uniform jet with sharp edges that expands
laterally at the sound speed in the jet comoving frame (model 2; solid
lines). {\it Right panel}: comparing model 2 ({\it inset}) to the
light curves calculated from a 2D hydrodynamical simulation of an
initially uniform jet with sharp edges.}
\label{fig:off-axis_LC}
\end{figure*}

This simple model of a point source along the jet symmetry axis
(called model 1 in Figure~\ref{fig:off-axis_LC}) is instructive, as it
captures much of the qualitative behavior of off-axis light curves.
Nevertheless, it is important to also consider more realistic jet
models. A useful model is a uniform jet of half-opening angle
$\theta_{\rm j}$, with an initial value $\theta_0$, which expands laterally
at some speed $\beta_{\rm s} c$ in its own, comoving rest frame (where
$\beta_{\rm s} = 0$ corresponds to no lateral expansion and $\beta_{\rm s}
\sim 1$ corresponds to relativistic expansion in the comoving
frame). For model 2 in Figure~\ref{fig:off-axis_LC}, $\beta_{\rm s}c$ is taken
to be the local sound speed, which implies $\beta_{\rm s}\approx 1/\sqrt{3}$ during the
relativistic stage. The dynamics are given by conservation of energy
and rest mass. The light curves for model 2 differ considerably from
those for model 1 for $\theta_{\rm obs} \lesssim \theta_0$ at early
times. This is because in model 2, for such viewing angles, the
observed flux at early times is dominated by material along the line
of sight, whose beaming cone encompasses the line of sight from the
very beginning, while in model 1 there is no emission along the line
of sight, and the emission from the point source along the jet axis is
strongly beamed away from the observer. Moreover, in model 2 the light
curves for $\theta_{\rm obs} \leq \theta_0$ are very similar to those
for $\theta_{\rm obs} = 0$, since the observed emission is mainly from
an angle of $1/\Gamma$ around the line of sight, and for such viewing
angles this region is mostly (or even totally) occupied by the uniform
jet, resulting in very small differences in the observed flux density.

We note that model 1 can be made somewhat more realistic by placing
the point source at the edge of the jet at the point P closest to the
line of sight for viewing angles outside of the jet
aperture, i.e., using $\theta = \max(0,\theta_{\rm obs}-\theta_{\rm j})$
instead of $\theta =\theta_{\rm obs}$ for the angle between the point
source and the line of sight. In model 2, for viewing angles
$1/\Gamma_0 \ll\theta_{\rm obs} - \theta_0 \ll \theta_0$ at early
times when $\Gamma(\theta_{\rm obs} - \theta_{\rm j}) \gg 1$ the emission is
dominated by the part of the jet that is closest to the line of
sight. Such emission arises from the area within the jet where the
Doppler factor is close to its highest value (at point P), which is
over a solid angle of the order of $\sim (\theta_{\rm
obs}-\theta_{\rm j})^2$. Therefore, if there were no lateral spreading,
the initial rise in the flux would be similar to that for a point
source at a fixed angle from the line of sight (which was discussed
earlier). If, however, the jet spreads laterally, $\theta_{\rm
obs}-\theta_{\rm j}$ would decrease in time, but the effect of the reduction
in the fraction of the jet contributing to the observed emission is
overwhelmed by the more rapid increase in flux due to the faster rate
at which the beaming cone of the emitting material approaches the line
of sight.

For the prompt emission, the ratio of $E_{\rm peak}$ -- the photon
energy at which the $\nu F_\nu$ spectrum peaks -- between lines of
sight outside the jet ($\theta_{\rm obs} > \theta_0$) and those within
the jet (which are similar to $\theta_{\rm obs} = 0$) is 
$a\approx 1/[1+\Gamma^2(\theta_{\rm obs}-\theta_0)^2] \sim
[\Gamma(\theta_{\rm obs}-\theta_0)]^{-2}$. The ratio of their apparent
isotropic equivalent energy, $E_{\rm\gamma,iso}$, which is the ratio
of their observed fluence, is $\sim a^2$ for viewing angles that are
relatively close to the edge of the jet, $1/\Gamma_0 < \theta_{\rm
obs}-\theta_0 < \theta_0$ (Eichler \& Levinson 2004), since in this
case only a fraction $\sim (\theta_{\rm obs}-\theta_0)^2/\theta_0^2
\propto 1/a$ of the jet (in the part closest to the line of sight)
contributes significantly to the observed emission, resulting is a
suppression by a factor of $\sim 1/a$ relative to the result of $a^3$
for a point source. For a roughly standard $E_{\rm peak}$ at viewing
angles within the jet, this reproduces the observed correlation
$E_{\rm peak} \propto E_{\rm\gamma,iso}^{1/2}$, reported by Amati et
al. (2002), Lloyd-Ronning \& Ramirez-Ruiz (2002), and subsequently
Lamb et~al.\ (2005) using data from {\it BeppoSAX}, BATSE and {\it
HETE-2}, respectively.~\footnote{This empirical correlation has
inspired many theoretical investigations arguing for its origin
despite the many debates on its observational validity. The reader is
referred to Gehrels et~al.\ (2009) for a review.} For larger viewing
angles, $\theta_{\rm obs}
\gtrsim 2\theta_0$, we have $E_{\rm\gamma,iso}(\theta_{\rm
obs})/E_{\rm\gamma,iso}(0) \sim (\Gamma\theta_0)^2[\Gamma(\theta_{\rm
obs}-\theta_0)]^{-6} \sim (\Gamma\theta_0)^{-4}(\theta_{\rm
obs}/\theta_0)^{-6}$, which scales as $a^3 \sim (\Gamma\theta_{\rm
obs})^{-6}$. This is similar to the expected change of
$E_{\rm\gamma,iso}$ with viewing angle for a point source, since for
such viewing angles all of the jet contributes similarly to the
observed emission, and it may be reasonably approximated as a point
source.

A more accurate description of the jet dynamics can be achieved with
hydrodynamic simulations. The resulting light curves for an initially
uniform jet with sharp edges, for different viewing angles, are shown
in the {\it right panel} of Figure~\ref{fig:off-axis_LC} (model 3).
The initially sharp edges of the jet quickly become smoother and the
jet becomes non-uniform, especially near the edges, were there is a
sharp decrease in the Lorentz factor and energy per solid angle, and
the velocity is not in the radial direction. This results in a much
larger contribution to the observed flux at early times for large
viewing angles outside the jet initial aperture, which in turn gives a
slower rise in flux compared to a perfectly uniform jet with sharp
edges (such as model 2).

\begin{figure*}
\begin{center}
\epsfxsize=10.0cm \epsfbox{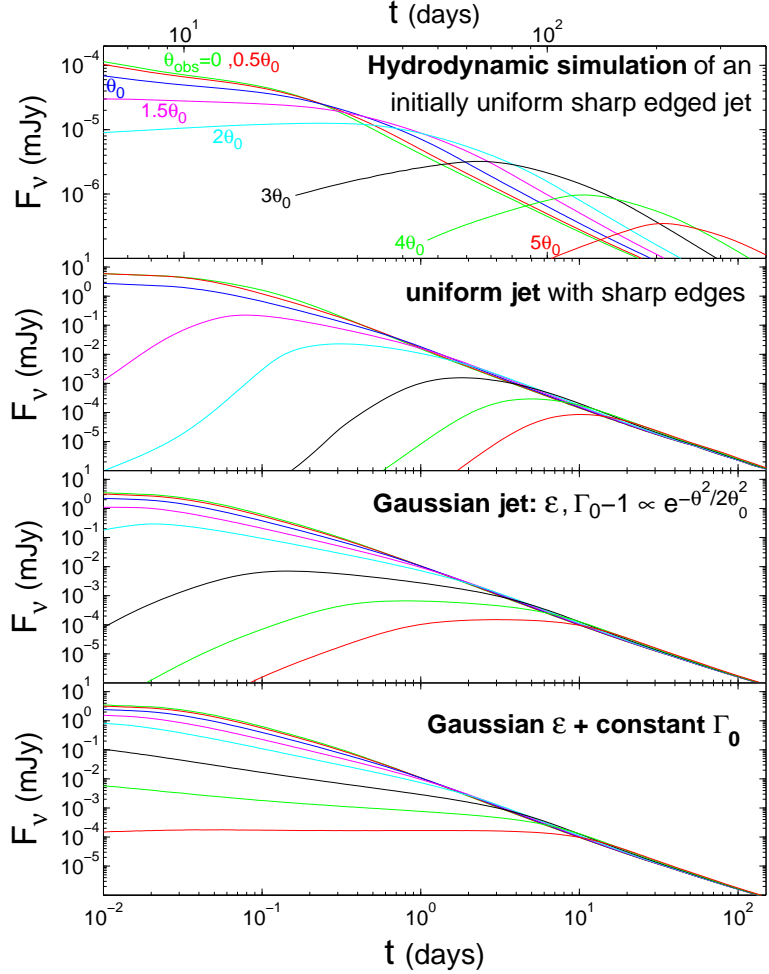}
\end{center}
\caption{Afterglow light curves for different jet structures,
dynamics, and viewing angles (from Eichler \& Granot 2006). The top
panel is from an initially uniform jet with sharp edges whose
evolution is calculated using a hydrodynamic simulation, the second
panel is for a uniform jet with sharp edges, and the two bottom panels
are for a Gaussian jet, in energy per solid angle, and either a
Gaussian or a uniform initial Lorentz factor. The viewing angles are
$\theta_{\rm obs}/\theta_0 = 0,\,0.5,\,1,\,1.5,\,2,\,3,\,4,\,5$, where
$\theta_0$ is the (initial) half-opening angle for the uniform jet
(two top panels) and the core angle ($\theta_c$) for the Gaussian jet
(two bottom panels).}
\label{fig:off-axis_LC4}
\end{figure*}

The light curves for different viewing angles depend not only on the
jet dynamics but also on its initial angular profile, and in
particular on $\Gamma_0(\theta)$ and $\mathcal{E}(\theta)$ (where
$\mathcal{E}$ is the jet energy per solid angle).
Figure~\ref{fig:off-axis_LC4} compares the light curves for different
viewing angles, between different jet structures. The first panel
from the top is for model 3 described above, the second panel is for a
uniform jet with no lateral expansion, while the last two panels are
for a Gaussian $\mathcal{E}(\theta)$, with either a Gaussian or a
constant $\Gamma_0(\theta)$. For a Gaussian jet, if both
$\mathcal{E}(\theta)$ and $\Gamma_0(\theta)$ have a Gaussian profile
(corresponding to a constant rest mass per solid angle in the
outflow), then the afterglow light curves are rather similar to those
for a uniform jet (Kumar \& Granot 2003). If, on the other hand,
$\mathcal{E}(\theta)$ is Gaussian while $\Gamma_0(\theta) = {\rm
const}$, then the light curves for off-axis viewing angles (i.e.,
outside the core of the jet) have a much higher flux at early times,
compared to a Gaussian $\Gamma_0(\theta)$ or a uniform jet (see the
bottom two panels of Figure~\ref{fig:off-axis_LC4}), due to a dominant
contribution from the emitting material along the line of sight which
has an early deceleration time in this case (Granot et~al.\ 2005).

\section{Unification schemes: implications of viewing angle effects}
\label{unification}

The appearance of GRBs depends so strongly on their orientation
relative to us that our current classification schemes might easily be
dominated by our random viewing angles rather than by more interesting
(intrinsic) physical properties. These inherently highly anisotropic GRB
models imply a radically different GRB appearance at different
viewing angles. In practice, GRBs of different orientations will thus
likely be assigned to different classes. Unification of these
fundamentally identical but apparently disparate classes is an
essential stepping stone to studying the underlying physical
properties of GRBs. The ultimate goal is to discover which are the
fundamentally important characteristics of GRBs -- e.g., black hole
mass, black hole spin, accretion rate -- and how they govern the
formation of jets, and the production of radiation.
In this section we thus critically examine GRB
off-axis models and contrast them with the afterglow observations of
X-ray flashes and sub-luminous long GRBs.  Since afterglow data in
these cases are too sparse and insufficient to derive meaningful
constraints on the overall population, we concentrate our efforts on a
few well-monitored examples.\\

\noindent
{\bf Empirical classification of GRBs:} GRBs traditionally have been
assigned to different classes based on their duration. On the basis of
this criterion, there are two classes of GRBs -- short and long --
dividing at $\sim 2\,$s duration (Kouveliotou et~al.\ 1993). GRBs have
also been classified according to their spectral properties, albeit
less successfully. In particular, bursts with lower photon energies
$E_{\rm peak}$ at which the $\nu F_\nu$ spectrum peaks have been
dubbed X-ray flashes (XRFs) based on observations by {\it BeppoSAX},
BATSE, and {\it HETE-2} (Heise et~al.\ 2001, Barraud et~al.\ 2003,
Kippen et~al.\ 2003, Lamb et~al.\ 2004, Sakamoto et~al.\ 2005). XRFs,
with durations ranging from several seconds to a few minutes and a
distribution on the sky consistent with being isotropic, are similar
to long duration ($\gtrsim 2\,$s) GRBs. In addition to XRFs, the
empirical classification of variable X-ray transients observed by {\it
HETE-2} had been expanded to include an intermediate class of events
known as X-ray rich GRBs (XRGRBs). The spectrum of XRGRBs and XRFs is
similar to that of long GRBs (Sakamoto et~al.\ 2005) except for their
lower values of $E_{\rm peak}$ and of $E_{\rm\gamma,iso}$ (their X-ray
and/or gamma-ray energy output assuming isotropic emission). In all
respects XRFs, XRGRBs and long GRBs seem to form a smooth continuum.

Many different models have been proposed for XRFs, most of which try
to incorporate them into a unified scenario with GRBs. These models
include high-redshift GRBs (Heise et~al.\ 2001), dirty (low-$\Gamma_0$)
fireballs (Dermer et~al.\ 1999, Huang et~al.\ 2002, Zhang et~al.\ 2003),
regular GRBs viewed off-axis (Yamazaki et~al.\ 2002, Dado et~al.\ 2004,
Kouveliotou et~al.\ 2004), photosphere-dominated emission (Drenkhahn
2002, Ramirez-Ruiz \& Lloyd-Ronning 2002, M\'esz\'aros et~al.\ 2002),
weak internal shocks (low variability, $\Delta\Gamma\ll\Gamma$; Zhang
\& M\'esz\'aros 2002, Barraud et~al.\ 2003, Mochkovitch et~al.\ 2004),
and large viewing angles in a structured (Lamb et~al.\ 2005) or 
quasi-universal (Zhang et~al.\ 2004) jet.  Most of these models mainly aim at
explaining the low values of $E_{\rm peak}$ in XRFs and do not address
their expected afterglow properties. The afterglow evolution alone
can, however, serve as a powerful test for XRF models. In fact, most
of the models discussed for XRFs and XRGRBs have at least one major
flaw in common: they do not naturally produce the very flat afterglow
light curve seen at early times. In what follows we thus concentrate
only on the class of models that naturally produce such light curves,
that is, a roughly uniform jet with sufficiently sharp edges viewed
from outside of the jet core. This class of models has been discussed
quantitatively in Sect.~\ref{off-axis_LC}.  Since most afterglow data of
XRFs and XRGRBs are too sparse, here we critically examine the role of
off-axis models and contrast them with afterglow observations of XRGRB
041006 and GRB\,031203 for which the afterglow light curves are
reasonably well monitored from sufficiently early times. The reader is
referred to {\it Swift} observations of XRF\,080330 (Guidorzi et
al. 2009) and GRB\,081028 (Margutti et~al.\ 2009) for more recent
examples of XRFs or GRBs, which bear similar characteristics to the
ones discussed here.\\

\begin{figure*}
\begin{center}
\epsfxsize=10.0cm \epsfbox{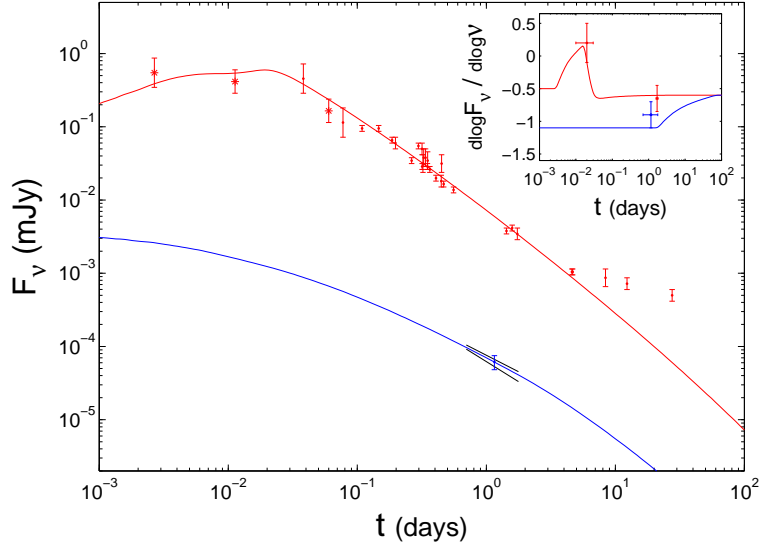}
\end{center}
\caption{A tentative fit to the optical R-band (upper curve) and X-ray   
($0.5-6\,$keV, lower curve) data of XRGRB\,041006 (see Granot et~al.\ 2005
and references therein).  The ROTSE-IIIa points are shown with
asterisk symbols since they are unfiltered, but they can still be
treated as R-band observations within the measurement errors. We also
added two lines to the X-ray data which indicate the edges of the $1\,\sigma$
confidence interval for the temporal decay index, $\alpha=1\pm 0.1$,
and cover the duration of the {\it Chandra} observation. The inset
shows the predicted spectral slope, $-\beta=d\log F_\nu/d\log\nu$, in
the optical (upper curve) and in the X-ray (lower curve), 
together with the values inferred from observations.}
\label{041006}
\end{figure*}

\noindent
{\bf Off axis jet models of XRGRBs and XRFs: The case of 041006.}
XRGRB\,041006 was detected by {\it HETE-2} (Galassi et~al.\ 2004). It
had a fluence of $5\times 10^{-6}\,{\rm erg\,cm^{-2}}$ in the
$2-30\,$keV range and $7\times 10^{-6}\,{\rm erg\,cm^{-2}}$ in the
$30-400\,$keV range, corresponding to $f_{X/\gamma}\approx -0.15$
which classifies it as an XRGRB. It has a redshift of $z=0.716$
(Fugazza et~al.\ 2004), which for a fluence of $f\approx 1.2
\times 10^{-5}\,{\rm erg\,cm^{-2}}$ in the $2-400\,$keV range gives
$E_{\rm\gamma,iso}\approx 1.6\times 10^{52}\,$erg. It had an observed
peak photon energy of $E_{\rm peak}^{\rm obs}=63^{+7}_{-5}\,$keV,
corresponding to $E_{\rm peak}^{\rm rest}=109_{-9}^{+12}\,$keV.
Figure~\ref{041006} shows an off-axis model yielding an acceptable fit
to the to the optical and X-ray afterglow observations of
XRGRB\,041006, which is also consistent with the upper limits at radio
and sub-mm wavelengths (Granot et~al.\
2005). From this analysis one can conclude that a successful
model for the afterglow of XRGRB\,041006 is that of a collimated,
misaligned jet interacting with a stellar wind external medium of mass
density $\rho_{\rm ext}=Ar^{-2}$, where $r$ is the distance from the
central source.  The parameter values used in this fit are:
$E=1.0\times 10^{51}\,$erg, $A_*\equiv A/(5\times 10^{11}\,{\rm
gr\,cm^{-1}})=0.03$, $\theta_0=3^\circ$, $\theta_{\rm obs} =
1.15\theta_0$, $p=2.2$, $\epsilon_e=0.1$, and $\epsilon_B=0.001$.

The optical light curve is very flat at early times ($\alpha\sim 0$ at
$t\lesssim 1\,$hr, where $F_\nu\propto t^{-\alpha}\nu^{-\beta}$) and
becomes steeper after a few hours ($\alpha\approx 1.2$), which is a
little steeper than the decay index in X rays at a similar time
($\alpha\approx 1$ at $t\approx 1\,$day). Also, the ratio of the flux
in optical and X rays at $t\approx 1\,$day implies a spectral index
of $\beta\approx 0.7-0.75$ assuming a single power law between them.
This suggests that the cooling break frequency $\nu_c$ is above the
optical after $1\,$day. Since one requires very extreme parameters to
get $\nu_c$ to the X-ray range after $1\,$day (even getting $\nu_c$ to
be above the optical after a day requires relatively low values of
$\epsilon_B$ and of the external density), it is most likely that
$\nu_c$ is between the optical and X-ray at $1\,$day, which can also
explain the steeper temporal decay index in the optical (by
$\Delta\alpha=0.25$) for a stellar wind environment ($k=2$). This
favors a wind medium over a uniform density one, since otherwise the
flux in the optical will decay more slowly than in the X-ray (also by
$\Delta\alpha=0.25$), which is contrary to what is observed for
XRGRB\,041006. At $t\gtrsim 5\,$days there is a flattening in the
optical light curve, which is probably due to an underlying SN
component or host. This explains why the observed flux is higher than that
predicted by our narrow relativistic jet model.

For a GRB jet with well-defined edges, both the prompt gamma-ray
fluence and the peak of the spectrum drop very sharply outside the
opening of the jet, as $a^2$ (or $a^3$ depending on $\theta_{\rm
obs}$) and $a$, respectively. Therefore, the low $E_{\rm\gamma,iso}$
of XRGRB\,041006 combined with $E_{\rm k,iso} = E/(1-\cos\theta_0)
\approx E(2/\theta_0^2)\approx 7.3 \times 10^{53}\,$erg (which serves
as a proxy for the $E_{\rm\gamma,iso}$ value that would have been
measured by an on-axis observer) implies $(E_{\rm
k,iso}/E_{\rm\gamma,iso})^{1/2}\sim 6.8$ and $\Gamma\sim(E_{\rm
k,iso}/E_{\rm\gamma,iso})^{1/4}(\theta_{\rm obs}-\theta_0)^{-1}\sim
330$. This implies a (cosmological) rest frame on-axis $E_{\rm
peak}(\theta_{\rm obs}<\theta_0) \sim (E_{\rm
k,iso}/E_{\rm\gamma,iso})^{1/2}E_{\rm peak} \sim 740\,$keV, which
falls closely within the observed $E_{\rm peak}-E_{\rm\gamma,iso}$
relation.\\

\noindent
{\bf Off-axis jet models for sub-luminous GRBs: the case of 031203.}
At a relatively small distance, with a redshift of $z = 0.1055$
(Prochaska et~al.\ 2004), GRB\,031203 was also atypical in its isotropic
equivalent gamma-ray energy output, with $E_{\gamma,{\rm iso}} \sim
10^{50}\,$ergs (Sazonov et~al.\ 2004).  In fact its $E_{\gamma,{\rm
iso}}$ was intermediate between GRB\,980425 and more typical bright
GRBs with $E_{\gamma,{\rm iso}} \sim 10^{52}-10^{54}\,$ergs (Frail et
al. 2001).  The gamma-ray light curve was smooth and similar to
GRB\,980425, consisting of a single peak lasting about $20\,$s and a
peak photon energy $E_{\rm peak} > 190\,$keV.  Soon afterwards, an
optical counterpart was identified and follow-up observations by
several telescopes revealed a supernova, SN~2003lw, with a spectrum
very similar to that of SN~1998bw (e.g., Malesani et~al.\ 2004).
Subsequent X-ray observations of GRB\,031203 with {\it XMM} and {\it
Chandra} identified an X-ray source coincident with the optical
transient. The flux decay rate and the isotropic luminosity of the
X-ray afterglow also ranked the event as intermediate between
GRB\,980425 and classical GRBs (Kouveliotou et~al.\ 2004).  A very faint
counterpart was also detected at centimeter wavelengths where it
displayed a peak luminosity more than two orders of magnitude fainter
than typical radio afterglows (Frail et~al.\ 2003), but again
comparable to that of GRB\,980425.

GRB\,031203, or at least its gamma-ray luminosity directed at us, was certainly
very weak. A straightforward interpretation might be that the GRB was
deficient in all its emissions in all directions.  This idea is
compatible with the afterglow light curve at radio
frequencies. However, when one combines the fact that a $20\,$s long
GRB was observed, as well as an X-ray and infrared afterglow, the
situation is more constrained and in fact more consistent with a
model in which GRB\,031203 was a typical powerful jetted GRB viewed
off-axis (Ramirez-Ruiz et~al.\ 2005).

\begin{figure*}
\begin{center}
\epsfxsize=10.0cm \epsfbox{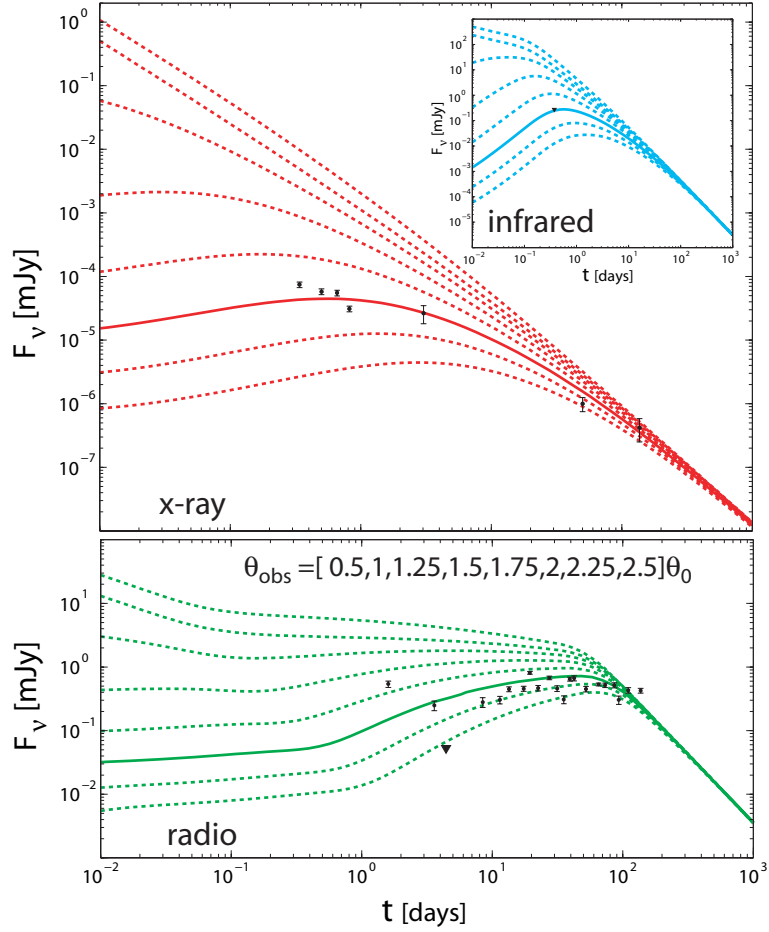}
\end{center}
\caption{
Afterglow emission from a sharp edged uniform jet in GRB\,031203. Light
curves are calculated for various viewing angles $\theta_{\rm obs}$ for a
GRB with the standard parameters $E_{\rm jet}=3 \times 10^{50}$ erg,
$p = 2.4$, $\epsilon_e = 0.15$, $\epsilon_B = 0.02$, $\theta_0 =
5^\circ$, and $A_*=(\dot{M}/10^{-5}\,{\rm M_\odot\,yr^{-1}})(v_w/
10^{3}\,{\rm km\,s^{-1}})^{-1}=0.1$. The data for GRB\,031203 can be
reasonably fit by different sets of model parameters (i.e., the
parameters cannot be uniquely determined by the data). For example, a
sharp-edged jet with $\theta_0 = 3.5^\circ$ seen at $\theta_{\rm
obs}\approx 2.25\theta_0$ gives also a reasonably good description of
the observations provided that $\epsilon_e = 0.1$ and $\epsilon_B =
0.04$.}
\label{031203}
\end{figure*}

The off-axis jet interpretation for GRB\,031203 requires the viewing
angle to have been $\theta_{\rm obs}\sim 2\theta_0$
(Figure~\ref{031203}).  A misaligned jet with a typical energy
expanding into a stellar wind with properties similar to those of
Wolf-Rayet stars is thus consistent with the observations, especially
with the slow initial decline rates seen in both the X-ray and radio
afterglow. One question that naturally arises is whether the observed
gamma-ray flux of GRB\,031203 can be explained within the framework of
this model.  The low $E_{\gamma,{\rm iso}}$ of GRB\,031203
implies~\footnote{This follows from the scaling of $E_{\gamma,{\rm
iso}}$ with $a$, here assumed to be $\propto a^3$} $\theta_0 =
3.8^\circ ({E_{\gamma,{\rm iso}}/ 10^{50}\, {\rm erg}})^{-1/8} (E_{\rm
jet}/{3\times 10^{50}\,{\rm erg}})^{1/8}$
$({\Gamma\Upsilon}/{50})^{-3/4}$, where $E_{\rm jet}$ is the kinetic
energy of the jet, and $\Upsilon=\theta_{\rm obs}/\theta_0-1$ and the
fiducial values were chosen to match those of GRB\,031203, which were
either observed ($E_{\gamma,{\rm iso}}\sim 10^{50}\,$erg) or inferred
from the fit to its afterglow ($\theta_0\sim 3^\circ-5^\circ$, $E_{\rm
jet}\sim 3\times 10^{50}\,$erg, $\Upsilon\approx 1$). A consistent
solution for both afterglow and prompt emission can thus be found if
$\Gamma\sim 50$ and $\Upsilon\approx 1$, which imply more typical
values of $E_{\rm p} \sim 2\,$MeV (given the observed value $E_{\rm
peak} \sim 190\,$keV) and $E_{\gamma,{\rm iso}} \sim 10^{53}\,$ergs
when observed on-axis (consistent within the intrinsic spread of the
$E_{\rm peak}-E_{\rm\gamma,iso}$ relation). These results are
applicable in the present context provided only that one further
condition is satisfied, namely, that the (on-axis) jetted outflow be
optically thin to high-energy photons.  For a burst with $E_{\rm peak}
\sim 2$ MeV, $\Gamma$ must exceed $\sim 50$. We thus conclude that the
observations, especially the slow initial decline rates seen in the
X-ray afterglow, are more consistent with an off-axis model in which
GRB\,031203 was a much more powerful GRB seen at an angle of about two
times the opening angle of the central jet.\\

\noindent
{\bf Unified schemes for GRBs:} The empirical classification scheme by
which an event is tagged as a GRB, sub-luminous GRB, XRGRB or XRF is
rather arbitrary. Therefore, there could be some cases where a jet
that is viewed on-axis ($\theta_{\rm obs}<\theta_0$) will be
classified as an XRGRB or XRF instead of as a GRB, or the opposite
case in which a jet viewed off-axis ($\theta_{\rm obs}>\theta_0$)
might be classified as a GRB instead of as an XRGRB or an XRF. A more
physically motivated classification would be according to the ratio of
the viewing angle $\theta_{\rm obs}$ and the jet half-opening angle
$\theta_0$, instead of relying purely on spectral characteristics as
in the present empirical scheme.  Such a classification would,
however, be much harder to implement as it is not a trivial task to
accurately determine the viewing angle. The strongest constraints
could thus be obtained from afterglow light curves of XRFs an XRGRBs
that are well monitored from early times and at various frequencies
(ranging from radio to X-rays). Current examples include {\it Swift}
observations of XRF\,080330 (Guidorzi et~al.\ 2009) and GRB\,081028
(Margutti et~al.\ 2009)

Similarly, the large statistical sample of GRBs and XRFs with redshift
measurements will allow a reconstruction of the intrinsic luminosity
function of the prompt emission.  If GRBs, XRGRBs and XRFs are only a
manifestation of the viewing angle for a structured, universal jet
(whose wings are producing the XRFs), then no break would be expected
in the luminosity function. On the other hand, if GRBs are the results
of viewing angles that intersect the jet (whether structured or not),
while XRFs and XRGRBs are off-axis events, then one would naturally
expect a break in the luminosity function. Guetta et~al.\ (2005) found
that a luminosity function with a break is favored in order for the
predicted rate of local bursts to be consistent with the observed
rate. This also prevents the existence of an exceedingly large number
of GRB remnants in the local Universe.

The relative fraction of XRFs and XRGRBs to GRBs is also expected to
be different in the various models. If indeed an XRF corresponds to
$\Gamma(\theta_{\rm obs}-\theta_0)\sim{\rm a\ few}$ and $(\theta_{\rm
obs}-\theta_0)\lesssim\theta_0$, then the solid angle from which an XRF
is seen scales as $\theta_0/\Gamma$ or as $\theta_0$ for a constant
$\Gamma$ (at a constant distance to the source), while the solid angle
from which a GRB is seen scales as $\theta_0^2$.  Therefore, the ratio
of solid angles for GRBs and XRFs scales as $\theta_0$, and more GRBs
compared to XRFs would be seen for larger $\theta_0$. As the distance
to the source increases, XRFs could be detected only out to a smaller
off-axis viewing angle, while most GRBs would still be bright enough
to be detected out to reasonably large redshifts. Therefore, the ratio
of GRBs to XRFs should increase with redshift. Finally, if the true
energy $E$ in the jet is roughly constant, then the maximal redshift
out to which a GRB could be detected would decrease with $\theta_0$
since $E_{\rm\gamma,iso}\propto\theta_0^{-2}$. This would increase the
statistical weight of narrow jets in an observed sample, as they could
be seen out to a larger volume.

Finally, we conclude with a few possible implications of the off-axis
model hypothesis for XRFs and XRGRBs.  For sufficiently large viewing
angles outside the edge of the jet, one might expect some decrease in
the variability of the prompt emission.  This is because the width
of an individual spike in the light curve scales as $\Delta t\propto
1/a$ while the peak photon energy and fluence scale as $a$ and $a^2 -
a^3$, respectively. Since the interval between neighboring spikes in
the light curve is typically comparable to the width of an individual
spike, $\Delta t$, then if $\Delta t$ increases significantly for
large viewing angles this would cause at least some overlap between
different pulses which would smear out some of the variability. Thus
one might expect XRFs to be somewhat less variable than GRBs, at least
on average, where a lower variability might be expected for lower
values of $E_{\rm peak}$ and $E_{\rm\gamma,iso}$. This may lead to a
simple physical interpretation of the observed variability-luminosity
relation in the prompt gamma-ray/X-ray emission (Fenimore \&
Ramirez-Ruiz 2000, Reichart et~al.\ 2001).

Another possible signature of the off-axis model for XRFs and XRGRBs
is in the reverse shock emission.  If the reverse shock is at least
mildly relativistic, then the optical flash emission would be less
beamed than the prompt X-ray or gamma-ray emission, due to the
deceleration of the ejecta by the passage of the reverse shock. This
might cause the optical flash to be suppressed by a smaller factor
relative to the gamma-ray emission, compared to the corresponding
on-axis fluxes. Thus XRFs or XRGRBs might still show reasonably bright
optical emission from the reverse shock, which might in some cases be
almost as bright as for classical GRBs. Finally, XRFs and XRGRBs might
also show a larger degree of polarization compared to GRBs.\\

Our understanding of GRBs has come a long way since their discovery
over 40 years ago, but these enigmatic sources continue to offer major
puzzles and challenges.  As we have described, our rationalization of
the principal physical considerations for the prompt and afterglow
radiation emanating from these objects combines some generally
accepted features with some more speculative ingredients. When
confronted with observations, it seems to accommodate their gross
features but fails to provide us with a fully predictive theory -- but
then again, no such theory exists as of yet. What is more valuable,
though considerably harder to achieve, is to refine models like the
ones advocated here to the point of making clearer quantitative
predictions, and to assemble, assess and interpret observations so as
to constrain or refute these theories.  What we can hope of our
present understanding is that it will assist us in such an endeavor.\\

\vspace{0.5cm}
\noindent {\bf Acknowledgments} J.G and E.R. gratefully acknowledge support from a Royal Society 
Wolfson Research Merit Award and a David and Lucile Packard Fellowship, respectively.

\end{document}